# Silicon Photonics WDM Transceiver with SOA and Semiconductor Mode-Locked Laser

Alvaro Moscoso-Mártir, Juliana Müller, Johannes Hauck, Nicolas Chimot, Rony Setter, Avner Badihi, Daniel E. Rasmussen, Alexandre Garreau, Mads Nielsen, Elmira Islamova, Sebastián Romero-García, Bin Shen, Anna Sandomirsky, Sylvie Rockman, Chao Li, Saeed Sharif Azadeh, Guo-Qiang Lo, Elad Mentovich, Florian Merget, François Lelarge, and Jeremy Witzens

*Abstract*—We demonstrate a complete Silicon Photonics WDM link relying on a single section semiconductor mode-locked laser and a single SOA to support up to 12 multiplexed channels with a bit error rate of 1e-12 at serial data rates of 14 Gbps without channel pre-emphasis, equalization or forward error correction. Individual channels reach error free operation at 25 Gbps and multi-channel operation at 25 Gbps is shown to be compatible with standard 7% overhead hard decision forward error correction. Silicon Photonics transmitter and receiver chips are hybridly integrated with driver and receiver electronics. A detailed link model is derived and verified. Particular emphasis is placed on accurate system level modeling of laser RIN, SOA amplified spontaneous emission noise and receiver noise. The impact of the electrical receiver bandwidth and non-Gaussian statistics on level dependent amplified spontaneous emission noise are investigated in detail. The channel count scalability as limited by SOA saturation is further analyzed taking cross gain modulation and four wave mixing into account. While semiconductor mode-locked lasers have been identified as a potential light source for low cost Datacom WDM transceivers for some time, this is, to the best of our knowledge, the first comprehensive investigation of the overall link budget in a Silicon Photonics implementation showing this technology to be a credible contender for low latency datacenter interconnects.

## I. INTRODUCTION

SINCE the first demonstration of a Silicon Photonics (SiP) Resonant Ring Modulator (RRM) in 2005 [1], a thorough understanding of the device's characteristics has been achieved [2], [3] enabling application specific optimization and reconfiguration [4]. Utilization of RRMs in high-speed Wavelength Division Multiplexed (WDM) interconnects in particular [5] appears a natural application leveraging their wavelength selectivity, compactness and high speed. Nevertheless, in order to demonstrate a fully integrated and fully functional WDM transceiver some challenges have to be addressed.

A number of options are available for the choice of an adequate light source. One possibility is to use an array of Distributed Feedback (DFB) lasers, each producing one of several optical carriers with high power and low Relative Intensity Noise (RIN), which are subsequently multiplexed into a single bus waveguide [6]. However, in this case each carrier needs to be independently controlled in order to maintain its spectral alignment to a fixed grid. Moreover, an on-chip multiplexer can result in significant additional Insertion Losses (IL) and fitting a large number of lasers into a single module might result in high cost. Another solution consists in using a semiconductor Mode-Locked Laser (MLL) producing a frequency comb with a fixed carrier spacing [7], [8]. This solution stands out by its compactness and conceptual simplicity, however the limited line power and increased RIN associated with MLLs has to be accommodated in the link budget. The line power is of particular importance in SiP since Silicon (Si) does not have a direct bandgap and, consequently, the light has to be coupled from a hybridly [9], [10] or heterogeneously [11], [12] integrated III-V semiconductor laser. Hybrid integration in particular, while enabling the integration of a state-of-the-art known good laser die, also results in additional optical interfaces and associated power losses. While significant progress has been made in the design of more efficient optical couplers [13]-[16], hybrid integration is still associated with significant optical losses that are compounded by further losses incurred in downstream SiP fiber-to-chip couplers. Therefore, when selecting a hybridly integrated MLL as a light source downstream amplification has to be considered. Since Erbium Doped Fiber Amplifiers (EDFA) are bulky, power hungry and expensive devices, Semiconductor Optical Amplifiers (SOA) are considered here as a possible solution offering sufficient gain and output power. However, nonlinear effects occurring in SOAs have then also to be taken into account in addition to excess noise resulting

This work was supported by the European Union's 7[th] Framework Program under Contracts 619591 (Project BIG PIPES) and 279770 (Frontiers of Integrated Silicon Nanophotonics in Telecommunications).

A. Moscoso-Mártir, J. Müller, J. Hauck, S. Romero-García, B. Shen, S. Sharif Azadeh, F. Merget and J. Witzens are with the Institute of Integrated Photonics (IPH) of RWTH Aachen University, Sommerfeldstr. 24, D-52074 Aachen, Germany (e-mail: jwitzens@iph.rwth-aachen.de). E. Islamova was with the Institute of Integrated Photonics and is now with Innolume GmbH.

F. Lelarge and A. Garreau are with III-V Lab, Campus de Polytechnique, 1 av. Augustin Fresnel, F-91767 Palaiseau Cedex, France (e-mail: francois.lelarge@3-5lab.fr). N. Chimot was with III-V Lab and is now with 3SPTechnologies.

R. Setter, A. Badihi, D. Rasmussen, M. Nielsen, A. Sandomirsky, S. Rockman and E. Mentovich are with Mellanox Technologies, Hakidma 26, Ofer Industrial Park, Yokneam, Israel (e-mail: mentovich@mellanox.com).

C. Li and G.-Q. Lo are with the Singapore Institute of Microelectronics (IME)/A*STAR, Science Park Road 11, Singapore Science Park II, Singapore 117685 (e-mail: logq@ime.a-star.edu.sg).

from Amplified Spontaneous Emission (ASE).

A number of recently demonstrated compact form factor SiP WDM systems have been designed for channel counts that are multiples of four [17], [18] in line with recent Multi-Source Agreements (MSAs) for Electro-Optic (E/O) datacenter transceivers, such as Quad Small Form-Factor Pluggable (QSFP) or C Form-Factor Pluggable (CFP) four channel modules, eight channel CFP2 modules or 12 channel CXP modules, with serial data rates up to 28 Gbps supporting Infiniband Extended Data Rate (EDR), 4-lane 100G Ethernet and 32G Fibre Channel.

On the Receiver (Rx) side both Germanium (Ge) photodiodes [19] and commercial grating coupled Flip-Chip Photodiodes [20] (FC-PD) have been used. Integrated Ge photodiodes have some advantages, such as i) having one fewer optical interface resulting in lower IL, as well as ii) having a lower capacitance due to their small size and a reduction of parasitics resulting from monolithic integration, that can be leveraged to increase the sensitivity of a co-designed Transimpedance Amplifier (TIA) [21]. However, Ge photodiode responsivity rapidly drops at wavelengths above ~1560 nm as determined by the direct bandgap of Ge. While the latter is narrowed by tensile strain resulting from the thermal coefficient mismatch with Si during post growth cooling of the material [22], thus extending the cutoff of the photodiode towards larger wavelengths, this cutoff remains close to the wavelength range of interest for the utilized MLL technology. On the other hand, FC-PD based on lower bandgap III-V materials can cover a wider wavelength range, for example 1260 to 1620 nm for typical InGaAs/InP photodiodes [23].

A number of options exist for integration with electronics, spanning wire bonding [19], [20], [24], flip-chip integration [17] and monolithic integration [25]. While monolithic integration and micro-bump bonding result in reduced parasitics and facilitate architectures such as distributed drivers [26], they are also less advantageous from the perspective of thermal management since all the heat sources are combined on a single chip increasing the thermal exposure of the laser.

In this paper we demonstrate an end-to-end solution for a SiP Datacom link combining a semiconductor single section MLL [27] with RRMs on the Transmitter (Tx) side. Both flip-chip and integrated Ge Waveguide Photodiodes (WPD) [28] are evaluated for the Rx. Modulator driver and TIA chips from Mellanox Technologies specifically developed for Datacom applications and verifying QSFP electrical signaling specifications are hybridly integrated via wire bonding. A complete SOA reamplified link is thoroughly characterized at 14 and 25 Gbps and a comprehensive link budget validated: The Tx and Rx subsystems are first individually investigated and the interoperability of the on-chip photonic components with 25 Gbps driver and Rx electronics verified. A detailed link budget is then derived and verified against the experimental data. In a first step, the Tx subsystem is characterized using a comb line from a semiconductor MLL as a light source combined with test equipment grade driver electronics. Due to limitations of the currently utilized test boards (no access to the thermal resonance frequency tuners of the RRMs since the corresponding pins were not connected) the SiP Tx with hybridly integrated electronics is characterized using a tunable bench top laser from Keysight technologies whose wavelength is adjusted to replace the RRM tuning functionality. While the combination of full electronic module integration with a MLL light source still remains to be performed in the future, the detailed characterization data reported here is sufficient for calibrating and verifying a predictive link model and extrapolating the capabilities of the technology, since co-operability of the SiP devices with the wire bonded electronics as well as operation of the photonic link with a MLL are separately verified and associated penalties extracted from the data.

The paper is divided into three parts: The first part focuses on the Tx, its constituting E/O devices, Tx characterization with a semiconductor MLL as a light source, and interoperability with wire bonded driver electronics. The second part of the paper treats the Rx integration and noise floor measurements. An experimental study of the effect of electrical filtering inside the Rx on level dependent ASE noise is of particular note (the detailed theory of which has been separately published in another publication [29]). Finally, in the third part the full link is experimentally characterized and a comprehensive link model verified against experiments. The channel count supported by the technology is predicted based on this data and on the detailed characterization of saturation effects in the commercial quantum well based SOA with which the system experiments have been performed (Thorlabs S9FC1004P).

## II. TRANSMITTER

The targeted Tx architecture (Fig. 1) consists in a MLL with a 100 GHz Free Spectral Range (FSR) coupled to a SiP chip. Since a downstream SOA is meant to be operated in the linear regime, after entering the chip unused comb lines are removed by a wideband filter so as to ensure the available SOA output power can be optimally allocated to the modulated lines. Filtering occurs before modulation since this reduces the requirements on the filter's spectral alignment and passband edge steepness. The remaining carriers are then modulated by a RRM array optimized for a drive voltage of 2 $V_{pp}$ (0 V to 2 V reverse bias). The electrical signal is generated by a driver from Mellanox Technologies with integrated Clock Data Recovery (CDR) and signal reshaping. While optional pre-emphasis of the optical channel is implemented in the driver, we have not been using it as the bandwidth of our RRMs is sufficient to sustain 25 Gbps signaling as is. Subsequently, all optical channels are jointly amplified by a SOA prior to being sent to the Rx. Amplification after modulation allows for compensation of the high attenuation induced by the RRMs operated close to resonance (low optical carrier to resonance frequency detuning) with a relatively low electrical signal level (see section II.A.2), while amplification prior to modulation would not help much due to the limited output power of the SOA. Furthermore, boosting the optical power levels prior to sending the light through the RRMs could lead to difficulties related to bistabilities and self-pulsation inside the rings that emerge as the optical power levels increase [30] and would be further exacerbated here by operating the RRMs close to resonance.

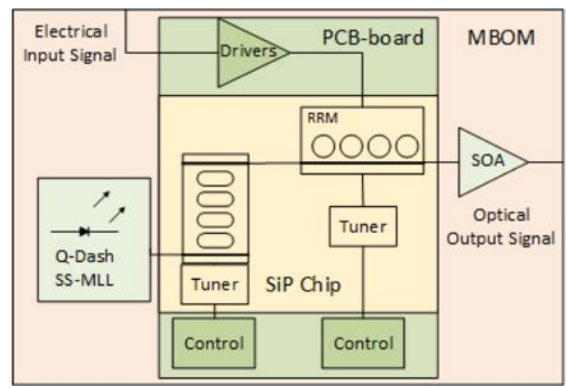

Fig. 1. Intended integrated Tx architecture: Mid-Board Optical Module (MBOM) with Printed Circuit Board (PCB) including drivers and SiP chip.

While the laser and SOA are intended to be hybridly integrated with the Tx chip in a compact module form factor, for the experimental evaluations of the link done here the wideband optical filter is implemented as a commercial off-chip device and the MLL, SOA and filter are fiber coupled to the Tx SiP chip. Ongoing work relating to MLL and SOA integration as well as to filter design will be reported in future publications. In the following, the key devices (MLL and RRM) are first described followed by a subsection on Tx characterization.

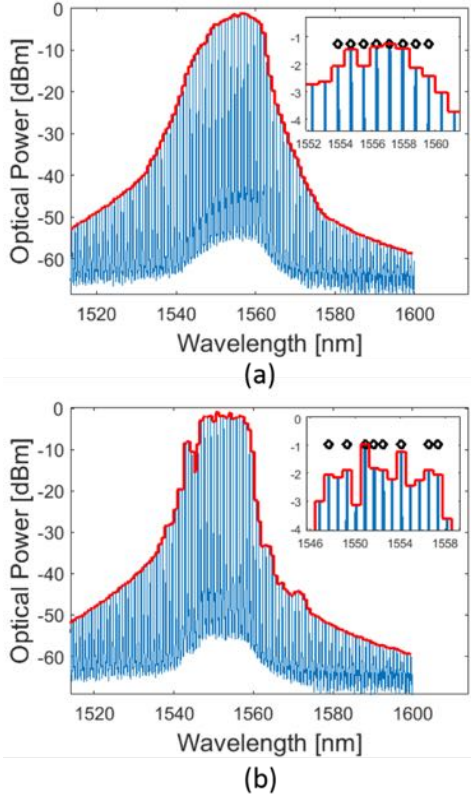

Fig. 2. Comparison between the optical spectrum of a single section MLL in a temperature and current range in which it is not mode locking, 300 mA and 30 °C (a), and a current and temperature range in which it is locking, 250 mA and 30 °C (b). The black markers in the insets represent the eight carriers with the highest line power. The laser diode gain material and laser stripe geometry are identical to the MLL reported in [31] but the length of the chip adjusted to obtain a 100 GHz FSR.

A. Transmitter Devices

1) Semiconductor Single Section Mode-Locked Laser

As already mentioned, choosing a passively mode-locked single section MLL as a WDM light source has the advantage of providing a compact solution with a fixed carrier grid, but also presents additional challenges in regards to RIN and operational stability. It is a well-known fact that the individual lines of a Fabry-Perot laser cannot serve as independent optical carriers due to mode partition noise resulting in excessive RIN: While the total power emitted by a Fabry-Perot is relatively stable (low RIN as measured over the entire spectrum), the optical power contained in individual lines undergoes strong fluctuations as the total power is dynamically reallocated between the lines. Mode partition noise is largely suppressed in MLLs by means of the mode locking. Recently, we have been able to measure RIN as low as -120 dBc/Hz decaying to its shot noise limit above 4 GHz on isolated comb lines of a single section MLL provided by III-V Lab [31]. A further characteristic of mode locking in single section MLLs is a flattening of the laser spectrum [32] which facilitates providing carriers for a substantial number of WDM channels with a well-defined optical power. Unfortunately, while the spectrum flattens overall over the entire center region of the comb as its shape changes from a bell shaped curve to a flattop distribution when entering an injection current / temperature range in which the laser mode-locks, the power variations between adjacent channels also become more pronounced and can exceed 1 dB. An example where this effect is relatively pronounced is shown in Fig. 2.

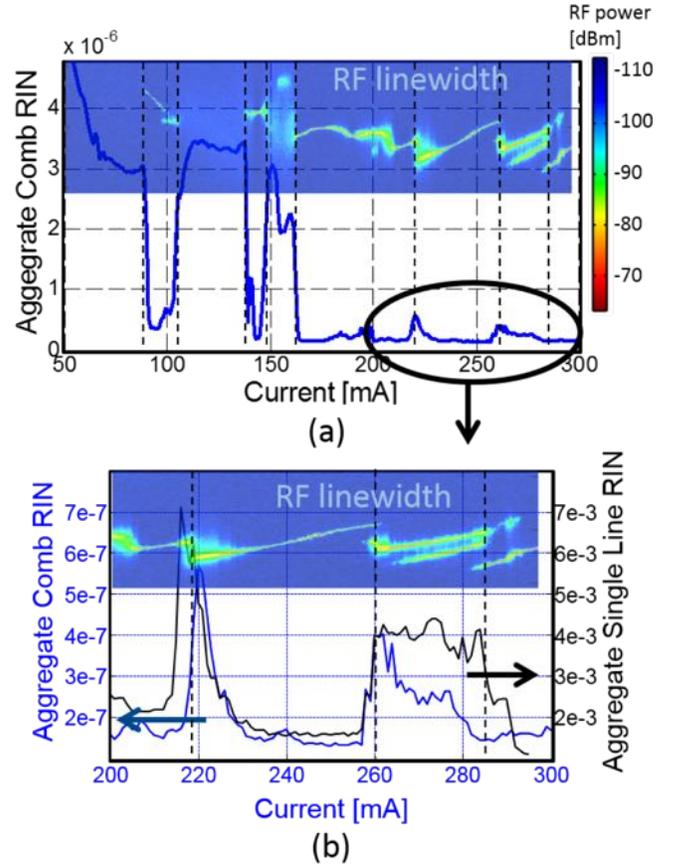

Fig. 3. (a) RIN of entire comb (blue curve) and spectrum of the RF beat note (color plot). The spectrum of the RF beat note is shown in the frequency range from 60.8 to 60.85 GHz (unlabeled y-axis of the color plot) for the MLL from [31]. The aggregate comb RIN (blue curve) was integrated from 8 to 300 MHz. (b) RIN of the entire comb (blue curve), RIN of an isolated comb line (black curve) and spectrum of the RF beat note (color plot, same frequency range as in (a)) in a zoomed in injection current range. The aggregate comb RIN was integrated from 8 to 300 MHz as in (a) and the single line RIN was integrated from 5 MHz to 20 GHz and represents the total RIN of the line since the RIN spectrum rolls off and reaches shot noise levels above 4 GHz. A clear correlation between mode locking (as evidenced by a single narrowband RF line), reduced comb RIN and reduced single line RIN is apparent, confirming mode locking to be a prerequisite for the usability of isolated comb lines as optical carriers. This data was taken at a laser operation temperature of 30 °C.

With the current state of the art of semiconductor MLLs, in order to obtain optimum locking performance prior to performing system level experiments (and prior to operating a transceiver module), the injection current and temperature set points have to be predetermined (Fig. 3). An outstanding challenge for future MLL designs is to reliably and repeatedly obtain mode locking in *predetermined* current and temperature ranges, a breakthrough that would also greatly facilitate taking such a technology to production. Integration of Distributed Bragg Reflectors (DBR) on chip [33], [34] provides additional degrees of design freedom such as tailoring the dispersion and width of the reflection frequency band [35] that may be conducive to reach this objective.

The Radio Frequency (RF) linewidth of the laser results from the beat note between adjacent comb lines and is a measure of the degree of correlation of their phase noise. A narrow RF linewidth is the primary indicator of mode locking and correlates with reduced single line RIN. Figure 3 shows the correlation between the RF linewidth and RIN for the laser reported in [31] (here and in the rest of the paper the integrated RIN is reported as $RIN_t = \sigma_P^2 / P_{AV}^2$ where $\sigma_P$ is the standard deviation (std) of the optical power and $P_{AV}$ is the average optical power). The first graph (a) compares the RF beat note spectrum (color plot) with the integrated RIN of the entire comb (blue line) for different injections currents. When the RF beat note is well defined and the corresponding linewidth small (visible as a clear yellow line in the color plot), the comb RIN is also low. On the other hand, regions with a smeared out beat note (poor or no mode locking), transitions between mode locking regimes (suddenly changing FSR for small current or temperature changes), or coexistence of two RF lines [31], [36] result in a high RIN. The next graph (b) shows both the comb RIN (RIN taken over the entire spectrum), as well as the RIN of an isolated comb line in a zoomed in injection current region. A correlation between all three characteristics – RF linewidth, comb RIN and single line RIN – is clearly visible.

The characteristics of the 100 GHz FSR laser chosen for the system experiments reported here [37] are shown in Fig. 4. It combines both relatively high line power and moderate RIN (if somewhat higher than for the work reported in [31]). The MLL is a Quantum Dash (Q-Dash) single section MLL developed and fabricated by III-V Lab. It is based on a Buried Ridge Stripe (BRS) Fabry Perot cavity with a ridge width of 1.25 μm whose gain material consists of six layers of InAs Q-Dashes in an InGaAsP barrier grown on an InP wafer. The rear facet of the laser is provided with a highly reflective thin film coating, while its front facet is as cleaved. After characterizing the laser at different temperatures and injection currents a good operating point was identified at 25°C and 238 mA that was then used for subsequent system measurements. At this operating point the laser has a FSR of 102.6 GHz and a center wavelength of 1542 nm. Fifteen consecutive lines have power levels between -1.3 and +1.7 dBm and the total power of the comb is 11 dBm, both measured after coupling to a lensed fiber followed by an isolator. A picture of the laser coupling setup, the laser spectrum and the RIN spectrum for an isolated central line can be seen in Fig. 4. A more systematic study of the RIN per line is reported in section II.B.1 (Fig. 9(b)).

*2) Resonant Ring Modulators*

The RRMs used for the system experiments have a radius of 10 μm, resulting in a FSR of 10 nm. In addition to the main bus waveguide, a drop waveguide with a low coupling coefficient serves as a tap that can be used to monitor the operating point of the modulator. Modulation is achieved via a phase shifter based on the plasma dispersion effect implemented as a reverse biased pin diode (with a series resistance of 53 Ω and a capacitance of 39 and 29 fF at respectively 0 V and 2 V reverse bias). More detailed information on the modulator design can be found in [31] (in which it is referred to as the "third category of device"). Two Tx chips are used for the system experiments, which turned out to have somewhat different characteristics due to fabrication variability: Modulators on the first chip were found to have a loaded resonator Quality Factor (Q-factor) of $Q_{load}$ = 5050 (RRM1), while RRMs on the second chip have a reduced $Q_{load}$ of 4300 due to a slightly higher waveguide coupling coefficient as well as higher waveguide losses (RRM2). As a consequence, RRMs from the first chip, that are also closer to the targeted design parameters, show a better performance. Their maximum Optical Modulation Amplitude (OMA), defined as $10log_{10}\left(\frac{P_1-P_0}{P_{in}}\right)$, with $P_1$ and $P_0$ the power of the 1- and 0-bit-states inside the bus waveguide right after the RRM and $P_{in}$ the power inside the bus waveguide right before the RRM, is -6.4 dB for a 2 $V_{pp}$ drive voltage and at the laser frequency to RRM resonance frequency detuning (concisely referred to as optical carrier detuning in the following) resulting in the highest OMA. Modulators from the second chip exhibit a reduced OMA of -7 dB. On the other hand the combined Grating Coupler (GC) and on-chip bus waveguide losses, respectively 10.3 dB and 9.7 dB for the two chips, are 0.6 dB better for the second chip. This information is provided upfront since it will be important for the comparison of individual test results in the following sections – the first chip was used for the Tx characterization with MLL and instrument grade bench top driver electronics while the second chip was wire bonded to chip scale Tx electronics and characterized with a bench top tunable laser. It should also be noted that GCs were not fully optimized in this chip iteration and we have improved them since from ~5 dB IL per GC to better than 3.5 dB in a later chip iteration fabricated in the same full flow process. These improved ILs are further reduced to 3 dB after permanent attachment of a fiber array with index matched epoxy, i.e., 4 dB of link budget improvement would be attainable with this improvement alone. This improved GC design is also used for the Rx chips reported in section III.

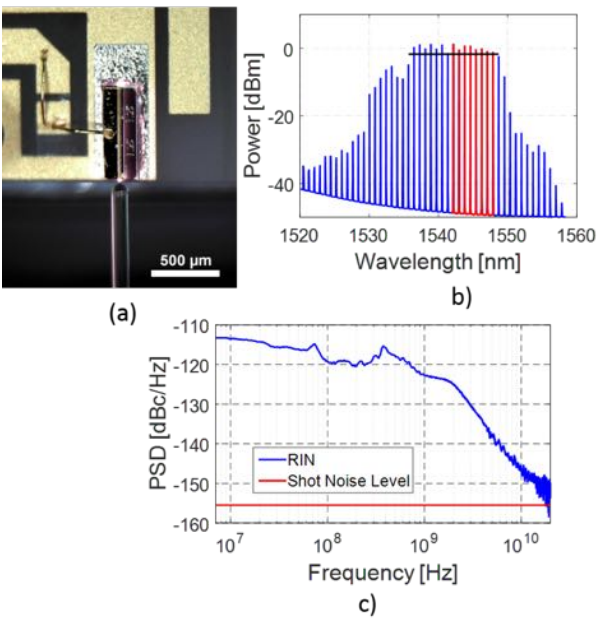

Fig. 4. MLL mounted on a ceramic submount and coupled to a lensed fiber (a), its recorded optical spectrum (b) and the RIN spectrum of one of the central lines (c). In (b) the black line marks the 15 lines within 3 dB of the peak line power. The lines for which signal Q-factors were measured in the system experiments reported below are shown in red. In (c) the RIN power spectral density of the line with center wavelength at 1546 nm is shown.

In the case of an unamplified system in which noise is typically dominated by additive Rx noise the optimum operating point of a modulator corresponds to the highest achievable OMA (assuming the resulting modulator cutoff frequency to also be sufficient, since the latter also depends on the operating point in the case of a RRM). On the other hand, in an amplified system the best operating point is between the points with the highest OMA and the highest extinction. Indeed, the signal-ASE beat noise resulting from ASE generated by an optical amplifier also depends on the signal level, so that full extinction is desirable to reduce the zero level noise, resulting in an additional performance metric in addition to OMA and bandwidth. Moreover, higher extinction also reduces the effect of 0-level RIN and allows increasing the channel count while maintaining optical power levels below SOA saturation. Higher extinction is achieved by reducing the optical carrier detuning at the cost of higher RRM IL (defined here as an on-chip metric as $10log_{10}\left(\frac{P_1}{P_{in}}\right)$), reduced OMA and reduced E/O bandwidth [3] as shown in Figs. 5 and 6, in which extinction, OMA and bandwidth are plotted versus detuning. While the OMA and extinction of RRM1 (Fig. 5(a)) were measured under DC-conditions, this was not possible for RRM2 (Fig. 5(b)) since it is wire bonded to an AC coupled modulator driver cutting off the signal at low frequencies. Thus its characteristics are obtained by analyzing eye diagrams resulting from modulation with a 4 Gbps Pseudorandom Bit Sequence (PRBS), which is quasi-DC given the high bandwidth of the Tx subsystem. At higher data rates the modulator cutoff frequency also has to be taken into account to find the optimum optical carrier detuning. The RRM bandwidth is lowest at resonance ($f_0/2Q$ where $f_0$ is the carrier frequency and $Q$ is the loaded Q-factor) and increases with detuning due to the emergence of peaking in the E/O $S_{21}$ [3] as can be seen in Fig. 6.

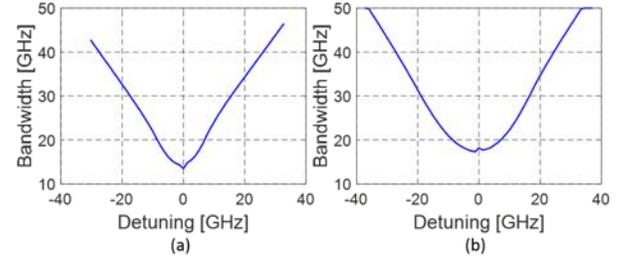

Fig. 6. RRM bandwidth vs. optical carrier detuning for chips 1 (a) and 2 (b). The cutoff frequency is defined as the point where the E/O $S_{21}$ is 3 dB below its maximum (more conservative metric than 3 dB below DC $S_{21}$).

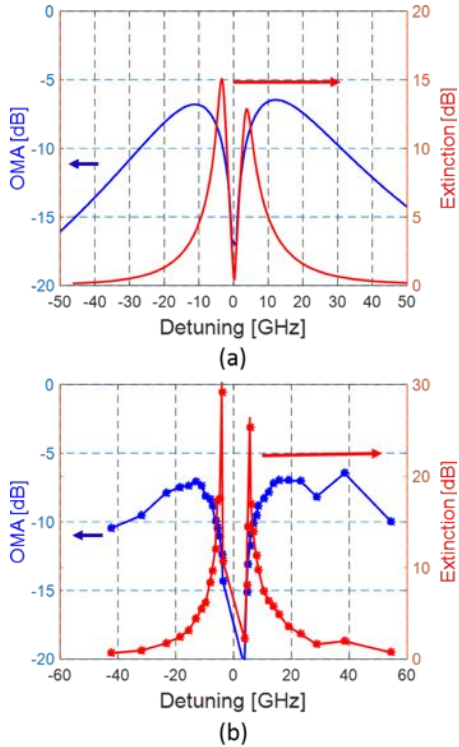

Fig. 5. RRM extinction and OMA vs. optical carrier detuning for chip 1 (a) and chip 2 (b). RRM characteristics are asymmetric in regards to positive and negative detuning due to dynamic free carrier induced absorption losses [3].

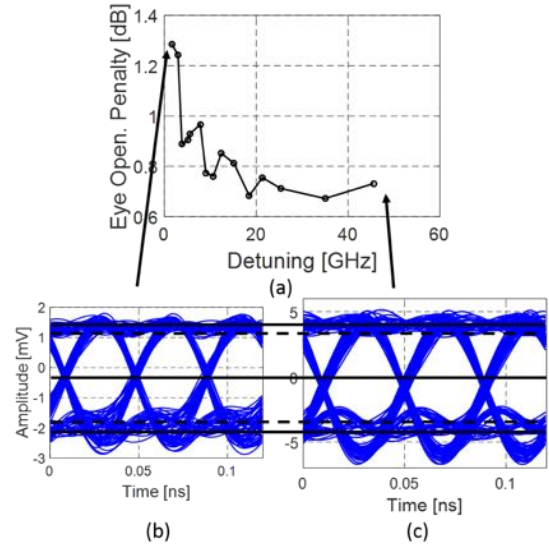

Fig. 7. (a) ISI penalty vs. optical carrier detuning extracted from eye diagrams at 25 Gbps for RRM2 (which is wire bonded to the modulator driver chip). (b) Eye diagram for low detuning and (c) eye diagram for high detuning. The eye diagrams have been rescaled for their DC 0- and 1-levels to be at the same height. The continuous horizontal black lines show the stabilized 0- and 1-levels as well as their mid-point. The dashed horizontal lines show the boundaries of the vertical eye opening of (c). It can be seen that the vertical eye opening relative to the DC levels (the ISI-penalty as defined in the text) is slightly worse for (b) and that the eye crossing is off-center in (c).

Figure 7(a) shows how the Inter-Symbol Interference (ISI) penalty extracted from eye diagrams depends on the optical carrier detuning, as measured for RRM2 at a data rate of 25 Gbps after wire bonding to the driver chip, wherein the ISI penalty is defined as the ratio between the vertical eye opening extracted from a signal averaged over repeated PRBS-7 cycles (to remove noise) and the OMA obtained with sequences of prolonged 0- and 1-bits, i.e, the DC OMA. At higher detuning, peaking in the RRM transfer function reopens the eye and reduces the (already quite small) ISI penalty (note that, as shown in [3], the signal distortion resulting from peaking is not always conducive to improve the eye opening, depending on the interplay with bandwidth limitations also occurring elsewhere in the link, here primarily the modulator driver to which RRM2 is wire bonded). The effect of the peaking can also be seen in the overshoot at the 0-levels as well as in the asymmetric eye crossing resulting from this overshoot (note that this electrical eye has been inverted relative to the optical eye by an inverting Rx, in the optical domain the overshoot is expected in the 1-level [5]). For carrier frequencies closer to resonance (Fig. 7(b)) the RRM bandwidth is reduced and therefore the ISI penalty increases.

Between the operating point at which RRM2 has the highest OMA (-7 dB at the larger detuning of 18 GHz) and a representative operating point at 7.5 GHz detuning, the estimated RRM cutoff frequency decreases from 32 GHz to ~20 GHz, the ISI penalty measured at 25 Gbps increases from 0.7 dB to 0.9 dB and the measured OMA drops to -8.4 dB, but the extinction also increases from 4.8 dB to 8.9 dB. This latter operating point is representative for the system level experiments described in section II.B.2 as it is close to the detuning resulting in optimum Tx + SOA optical output signal quality as quantified in the following by the signal Q-factor. The ISI penalty measured for RRM2 also includes the effect of driver bandwidth since it has been measured on the Tx subsystem. It should also be noted here that since the 0-level ASE noise is particularly sensitive to extinction, getting as close as possible to critically coupled RRMs is much more important here than in an unamplified optical link. A detailed modeling of the RRM characteristics as well as of the thermal tuners used to control their resonance frequency can be found in [31].

It should also be noted that in all the system experiments reported in this paper we work with positive optical carrier detunings, i.e., the frequency of the laser is above the resonance frequency of the RRM. The reason is twofold: First, this results in a slightly higher OMA, as the contributions to the OMA of dynamic waveguide losses occurring inside the RRM as the free carrier density is modulated and of the refractive index change stack up with the same sign [3]. The other reason is that self-heating induced bistability occurs at negative optical carrier detuning [38], thus limiting the maximum optical input power to the RRM in the absence of a control system or a fast offset compensation in the Rx (even then, dynamic suppression of these instabilities might result in a challenging control problem).

### B. Transmitter Subsystems

#### 1) Characterization with MLL Light Source and Test Equipment Grade Electronics

Figure 8 shows the test setup used to characterize the Tx chip using a MLL as light source. After coupling the MLL to a lensed fiber and sending it through an isolator, four comb lines are selected by a commercial passband filter with a tunable center frequency (1.4 dB IL). Note that since an eight or twelve channel filter was not available at the time of the experiments, the effect of channel count on data distortion induced by a saturating SOA cannot be fully accounted for here. The channel count scalability is independently investigated in section IV of this paper – this section is rather focused on a linear operation of the optical part of the link.

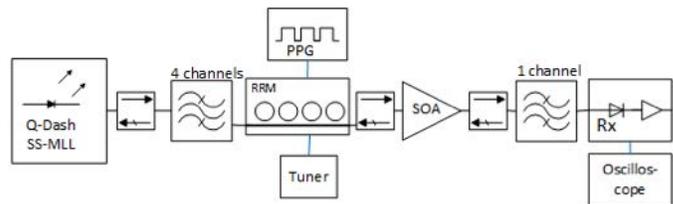

Fig. 8. Test setup including the MLL, RRM SiP chip and commercial SOA.

The light is subsequently grating coupled to a bus waveguide on a SiP chip comprising a RRM array (chip 1). After outcoupling from the SiP chip with a second GC, the light is sent to a commercial single polarization quantum well SOA interleaved between two additional isolators (24 dB gain, 15 dB output saturation power at the -3 dB gain compression point, 10 dBe Noise Figure (NF), each measured with the SOA connected to the two up- and downstream isolators and operated with a 650 mA drive current at 25 °C[1]). To emulate the SiP Rx, a commercial 40 GHz optical bandpass filter with a tunable center frequency (2.4 dB IL) is used to isolate a single modulated line that is then sent to a commercial 40 GHz bandwidth photoreceiver (Finisar/U2T XPRV2021A) with an input referred TIA noise density specified to be below $40\ pA/\sqrt{Hz}$ and a photodiode DC responsivity specified to be between 0.5 and 0.75 A/W. Finally, eye diagrams are visualized on a 20 GHz real time oscilloscope with a 4th order Bessel filter transfer function. The signal Q-factor ($Q_{sig}$), the std of the 0- and 1-level noise, the eye height and the ISI penalty are extracted by post processing the complete signal traces. Since a large number of complete PRBS-7 cycles are recorded (respectively 120 and 213 cycles at 14 and 25 Gbps), averaging over multiple cycles allows separating ISI from noise during the analysis (the complete methodology is described in [31]).

---

[1] At 650 mA current, the bare SOA die is specified by Thorlabs as having 30 dB gain, a NF of 6 dBe and a saturation power of 18 dBm, consistent with 3 dB chip to fiber losses at both ports, including the ILs of the isolators.

These measurements are repeated for eight consecutive MLL lines used as optical carriers in the system experiments described here. For these comb lines Fig. 9 gives an overview of line power (a), aggregate single line RIN as integrated between 5 MHz and 20 GHz (b), as well as the measured $Q_{sig}$ (c) obtained after modulation with a RRM from chip 1 with a 2 $V_{pp}$ drive signal at data rates of 14 and 25 Gbps (setup shown in Fig. 8). As can be seen all eight carriers achieve signal Q-factors of more than seven when modulated at 14 Gbps, the criterion required to achieve an uncorrected Bit Error Rate (BER) below 1e-12. However, in its current configuration the system seems to run into its performance limit at 25 Gbps as only one carrier achieves a $Q_{sig}$ above 7. Improvements required to enable reliable multi-channel operation at 25 Gbps are discussed in section IV.B.2 "Multi-Channel Transceiver Link Budget".

When comparing Figs. 9(a) and 9(b) with 9(c) a strong correlation between line power and RIN on the one hand and $Q_{sig}$ on the other hand can be seen. As expected, the best result was achieved for the line with highest power and lowest RIN.

In order to model the link budget, the dependency of $Q_{sig}$ on power and noise is analytically modeled. $Q_{sig}$ is defined as

$$Q_{sig} = \frac{\eta(P_{Rx,1}-P_{Rx,0})}{\sigma_1+\sigma_0} \quad (1)$$

where $P_{Rx,1}$ and $P_{Rx,0}$ are respectively the power levels of the logical 1- and 0-states at the input port of the Rx (for long strings of repeating 1s or 0s), $\eta$ is the eye opening penalty due to ISI and $\sigma_1$ and $\sigma_0$ are the std of the level dependent noise for the logical 1- and 0-states. The main noise sources are RIN, ASE and Rx noise, therefore the std are modeled as

$$\sigma_{0/1} = \sqrt{RIN_t P_{Rx,0/1}^2 + \sigma_{ASE,0/1}^2 + P_{n,Rx}^2 \Delta f_{Rx}} \quad (2)$$

where $\sigma_{ASE,0/1}$ is the std of the ASE noise for the 1- and 0-levels, $P_{n,Rx}$ is the input referred Rx noise spectral density (in $W/\sqrt{Hz}$), $\Delta f_{Rx}$ is the Noise Equivalent Bandwidth (NEB) of the Rx (here the oscilloscope's bandwidth since its transfer function is close to being square shaped) and $RIN_t$ is the aggregate RIN per line calculated as

$$RIN_t = \int_{f=0}^{f=\Delta f_{Rx}} 10^{\frac{RIN\left[\frac{dBc}{Hz}\right]}{10}} df \quad (3)$$

with $RIN$ being the measured power spectral density of the noise expressed in dBc/Hz. Since the RIN decays to its shot noise limit above 4 GHz, $RIN_t$ is insensitive to $\Delta f_{Rx}$ in the range compatible with the data rates investigated here and can be taken directly from Fig. 9(b).

In order to accurately calculate the std of the ASE noise, it has to be taken into account that level dependent ASE noise at the output of the optical amplifier is modified by the optical filter interposed between the SOA and the Rx as well as by the electrical low pass filtering occurring during Opto-Electronic (O/E) conversion or inside the Rx. We have derived analytic expressions accounting for electrical filtering in the level dependent ASE noise whose detailed derivation has been published in [29], so that here we restrict ourselves to report the expressions directly applicable to the experiments conducted here. Assuming both the electrical and optical filters to have ideal square shaped transfer functions with the electrical filter cutoff frequency denoted as previously as $\Delta f_{Rx}$ and the width of the optical filter passband denoted as $2\Delta f_{OF}$ (wherein $\Delta f_{OF}$ is the equivalent filter cutoff as applied to the single sided signal spectrum) and assuming the data to be applied to the optical carrier prior to optical amplification, as is the case in our system architecture, we derive the 0- and 1-bit ASE-signal beat noise levels referred back to the input port of the Rx after electrical filtering to be

$$\sigma_{ASE,0/1} = \sqrt{N_n \left[ \begin{array}{c} P_{SOA,0/1} \min(\Delta f_{Rx}, \Delta f_{OF}) \pm \frac{P_{SOA,1}-P_{SOA,0}}{4} \mu f_N \\ -\gamma \frac{(\sqrt{P_{SOA,1}}-\sqrt{P_{SOA,0}})^2}{8} f_N \end{array} \right]} \quad (4)$$

with

$$N_n = 2G^2 F h f_0 \quad (5)$$

where $G$ is the effective optical gain between the input port of the SOA and the input port of the Rx (i.e., also taking optical losses between the SOA and the Rx into account), $F$ is the noise factor of the SOA, $h$ is Planck's constant, $f_0$ is the optical carrier frequency, $P_{SOA,0}$ and $P_{SOA,1}$ are the optical power levels entering the SOA for the logical 0- and 1-bits, and $f_N$ is the Nyquist frequency (half the signaling rate). $\mu$ and $\gamma$ are coefficients that depend on the exact signal shape (e.g., fall and rise times) and signal power spectral density, as well as on the values of $\Delta f_{Rx}$ and $\Delta f_{OF}$ relative to $f_N$ and relative to each other. Table I summarizes the numerical values of $\mu$ and $\gamma$ for the different situations encountered during Tx characterization, as evaluated with a numerical model taking the exact shape of experimentally recorded signal traces into account.

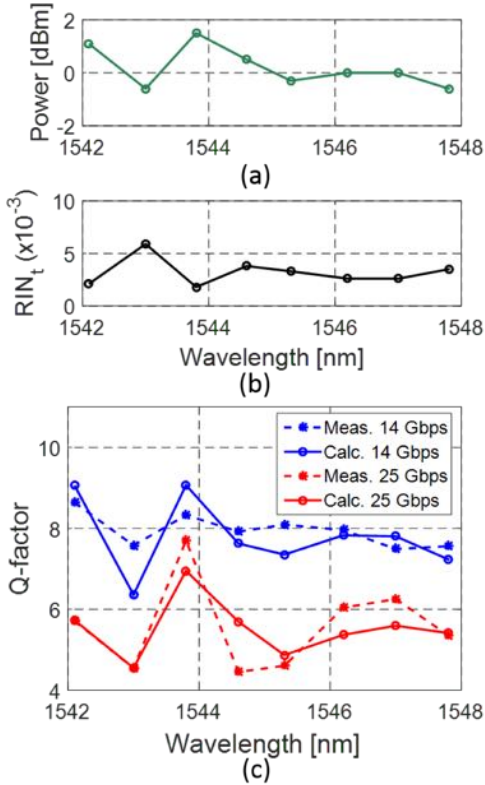

Fig. 9. (a) Power and (b) integrated RIN for the eight consecutive MLL lines used for the Tx characterization. (c) Measured and calculated $Q_{sig}$ for these eight carriers after modulation with a RRM from chip 1 with a 2 $V_{pp}$ drive signal. The RIN shown in (b) was integrated between 5 MHz and 20 GHz.

TABLE I
PARAMETERS FOR FILTERED LEVEL DEPENDENT ASE NOISE

| Data Rate | Optical Filter $2\Delta f_{OF}$ | Electrical Filter $\Delta f_{Rx}$ | $\mu$ | $\gamma$ |
|---|---|---|---|---|
| 14 Gbps | 40 GHz | 20 GHz | 0.51 | 0.27 |
| 14 Gbps | 40 GHz | 32 GHz | -0.06 | -0.16 |
| 25 Gbps | 40 GHz | 20 GHz | 0.63 | 0.88 |
| 25 Gbps | 40 GHz | 32 GHz | 0.06 | 0.10 |

While the same optical filter was used after the SOA in all the experiments ($2\Delta f_{OF}$ = 40 GHz), we have to distinguish between the electrical bandwidth of the real time oscilloscope used to record the eye diagrams (20 GHz) and the bandwidth of the analog front end of the Error Detector (ED) of the Bit Error Rate Tester (BERT) used in the following sections (32 GHz). In particular, when the electrical bandwidth of the Rx exceeds $\Delta f_{OF} + f_{max}$, i.e., the sum of half the optical filter passband and the maximum signal frequency $f_{max}$ (on the order of two times the Nyquist frequency), both $\mu$ and $\gamma$ are zero [29] and Eq. (4) reduces to the simple expression of ASE-signal beat noise for slowly varying signal levels. Given the high bandwidth of the ED, this condition is almost verified when recording BER at 14 Gbps. Even at 25 Gbps the numerical values of $\mu$ and $\gamma$ are much reduced when receiving the signal with the BERT's ED rather than with the 20 GHz real time oscilloscope.

Importantly, the net effect of Eq. (4) is to reduce the signal Q-factor: due to the downward concave convexity of the square root function the term $\pm(P_{SOA,1} - P_{SOA,0})\mu f_N/4$ increases $\sigma_{ASE,0}$ more than it decreases $\sigma_{ASE,1}$. Thus the differing electrical bandwidth of these two Rx configurations can induce (modest) discrepancies in the measurement data that will be further discussed in the following.[2]

In order to verify this model, we modulated a -14 dBm optical carrier with a commercial Mach-Zehnder Modulator (MZM) and amplified it with the SOA downstream of the modulator followed by the 40 GHz passband single channel optical filter. After being converted back to the electrical domain with the commercial U2T/Finisar Rx, the signal was recorded for data rates between 4 and 32 Gbps by the real time oscilloscope with the analog bandwidth set to 21 GHz. The std of the 1- and 0-levels was extracted and is shown in Fig. 10 (dots). We then calculated the expected std taking the effect of electrical filtering on the level dependent ASE-signal beat noise as predicted by Eq. (4) into account (continuous line). It is clearly apparent that Eq. (4) adequately reflects the experimental data and that with increasing data rates the effect of electrical filtering on ASE noise levels becomes more pronounced and has to be taken into consideration in order to accurately predict the signal Q-factor. An inflection in the curve showing the 0-level ASE noise above 20 Gbps can also be modeled by Eq. (4) once the reduction of the signal extinction due to optical filtering at high data rates is taken into account (ISI occurring due to the optical filter actually has to be referred back to the SOA input prior to applying Eq. (4) [29]).

---

[2] In previous publications [37] we had to empirically modify the values of $P_{SOA,0}$ and $P_{SOA,1}$ in order to take this effect into account in link models. This is no longer necessary in the signal Q-factor modeling done here.

Modeling the effective $Q_{sig}$ of the RRM + MLL Tx based on Eqs. (1)-(5) under consideration of the 20 GHz oscilloscope bandwidth ($\Delta f_{Rx}$) and the 40 GHz optical filter bandwidth ($2\Delta f_{OF}$), we obtain the curves plotted with continuous lines in Fig. 9(c). In addition to ASE, RIN and Rx noise are also taken into account (due to its restriction to low frequency components below ~4 GHz, level dependent RIN is not significantly modified by electrical filtering [29]). The input referred power spectral density of the Rx noise was independently measured to be 45 $pW \cdot Hz^{-0.5}$ (consistent with the specifications of the commercial Rx) and the RIN was set according to the data shown in Fig. 9(b). Optical carrier detunings were independently optimized for 14 and 25 Gbps operation in order to obtain the best possible $Q_{sig}$ and resulted in measured extinctions of respectively 11 dB and 6.5 dB and measured on-chip OMAs of respectively -8.4 (IL = 8 dB) and -6.9 dB (IL = 5.8 dB). The corresponding optical carrier detunings are estimated as 5.1 GHz and 8.1 GHz and the corresponding RRM E/O cutoff frequencies are estimated as 18 GHz and 22 GHz. The shifting of the optimum operating point towards a higher detuning (with lower extinction) at 25 Gbps is due to the need for more bandwidth. This resulted, at 25 Gbps, in measured ISI penalties between 0.9 and 1 dB.

Although the resulting Q-factors follow the trend of the measurements quite well and result in the correct average Q-factor values, we see some discrepancies. These can be partially due to small variations of the RIN as a consequence of changed laser feedback resulting from slight displacements of the lensed fiber, but they are mainly attributed to a rather rough estimate of SOA input power based here on the independent characterization data shown in Fig. 9(a) rather than on an inline measurement. Overall the model appears to be a good predictor for link budget calculations. Based on this model we predict the maximum acceptable RIN as a function of the MLL line power for the link architecture described above (shown in Fig. 8). The results are shown in Fig. 11 for 14 (red line) and 25 Gbps (blue line). The markers in the graph represent the 15 central lines of the MLL with power levels between -1.2 and 1.7 dBm and are colored based on whether a Q-factor of more than 7 was achieved at 25 Gbps (blue square) or only at 14 Gbps (red circles). The black crosses show the RIN and line power of the lines whose resulting $Q_{sig}$ was not measured. Our model appears to give a good prediction: While being a bit conservative (one line predicted to fail at 14 Gbps actually features a Q-factor above 7), the actual discrepancies between the model and the measurements remain modest (see Fig. 9(c)). Improving the cumulative GC losses by 4 dB, as already achieved in this process line, would already allow 5 lines of this MLL to reach a signal Q-factor above 7 at 25 Gbps (2 lines have almost identical power and RIN and cannot be distinguished in the graph).

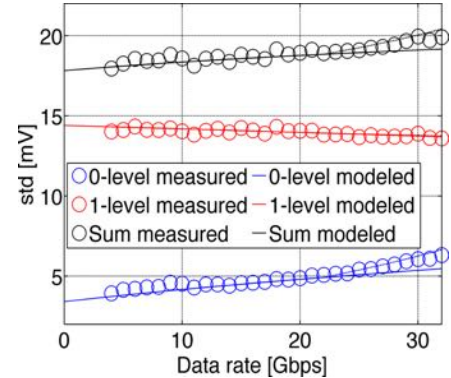

Fig. 10. Verification of the model described by Eq. (4). The dots show the measured values of the 0- and 1-level noise std, as well as their sum (the denominator of the signal Q-factor). The continuous lines correspond to the model given by Eq. (4). The doted lines take into account a progressive reduction of the signal extinction from 13 dB to 10 dB occurring for data rates between 20 and 32 Gbps due to signal clipping by the optical filter as well as the finite cutoff frequency of the optical modulator (this bandwidth limitation is also seen in the ISI penalty extracted from the eye diagrams). The reduced extinction further penalizes the 0-level ASE-signal beat noise.

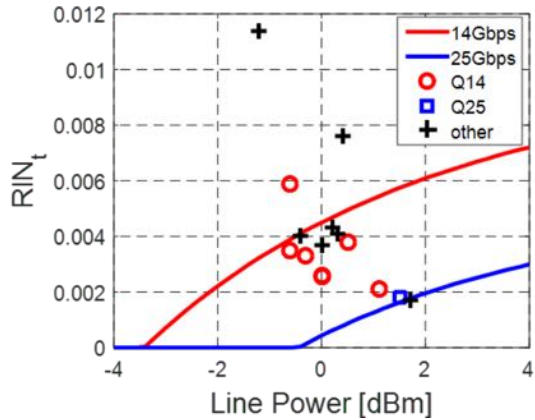

Fig. 11. Prediction of laser characteristics required to achieve error free (BER < 1e-12) 14 and 25 Gbps modulation for the link architecture shown in Fig. 8. Blue and red markers respectively represent the 8 consecutive comb lines with which 25 and 14 Gbps error free operation has been achieved. Black markers correspond to lines for which BER has not been measured.

*2) Transmitter Characterization with Hybridly Co-Integrated Driver*

After characterizing the Tx with test equipment grade electronics connected to the SiP chip via high-speed RF probe tips, we proceed with the integration of the E/O modulators with a driver chip from Mellanox technologies (part number MTxS28nn). A second Tx chip (RRM2) was mounted on an evaluation board and a single RRM wire bonded to the driver chip (Fig. 12) with two wire bonds transporting respectively ground and signal. The distance between corresponding pads is 300 µm, the diameter of the wire bonds 25 µm and the distance between the wire bonds 100 µm. The driver was designed to support signaling rates of up to 28 Gbd and to drive Franz-Keldysh direct absorption modulators with a comparable capacitance to our RRMs [39] and has a low output impedance of 4 Ω. It consists of a self-biased differential input stage, a programmable input equalizer, a Limiting Amplifier (LA), an optional CDR for high data rates, and a configurable driver output stage (set to source a 2 $V_{pp}$ single ended signal driving the modulator between 0 and 2 V reverse bias). Pre-emphasis of the E/O channel was not needed and disabled, since the bandwidths of the E/O components is sufficient as is to sustain a 25 Gbps link.

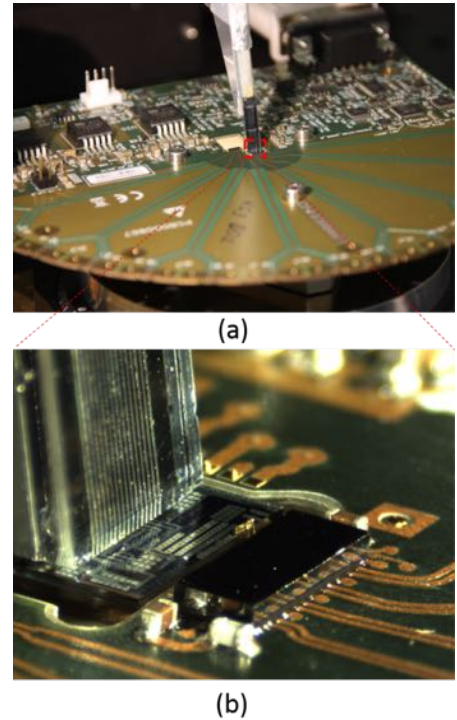

Fig. 12. (a) Test board with fiber array and (b) zoom on the SiP chip wire bonded to the modulator driver.

The test setup that we used to characterize the RRM integrated with the driver chip is depicted in Fig. 13. We used a Programmable Pattern Generator (PPG) from Anritsu (MU183020A 28G/32G bit/s PPG with 20% to 80% rise and fall times of 12 ps) to generate a PRBS-7 data stream routed to the differential input stage of the driver. Since the pins for accessing the thermal tuners of the RRMs were not routed off chip in this evaluation board, we used a bench-top tunable laser from Keysight Technologies (81600B) as a low noise light source instead of the MLL. Tuning of the carrier frequency was then used instead of the thermal RRM tuner to optimize the optical carrier detuning. While the higher RIN of the MLL comb lines is not taken into account in this set of experiments, interoperability between the RRMs and the driver as well as additional ISI resulting from parasitics associated with the hybrid driver integration are verified and characterized here. A full Tx model can then be derived by combining with the results of the previous section. As the tunable single mode laser only generates a single carrier, the 4-channel passband filter shown in Fig. 8 is not needed and is removed from the setup. As previously, the signal is amplified with the commercial SOA after modulation. A standard Rx is emulated with the Finisar/U2T Rx connected to either the real time oscilloscope (for signal Q-factor measurements) or the 32 Gbps ED from the Anritsu BERT (MU183040A 28G/32Gbit/s ED).

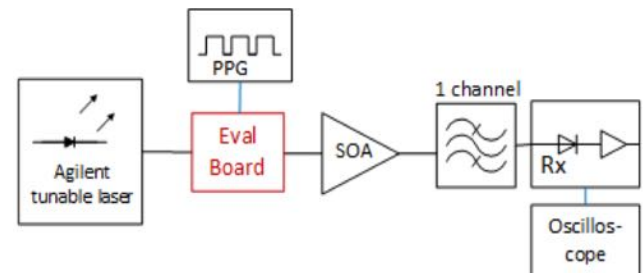

Fig. 13. Test setup including the evaluation board with RRMs wire bonded to the modulator driver chip and commercial SOA. As previously, isolators are placed before and after the SOA.

Figure 14(a) shows the measured Q-factors (blue dots) and BER (red line) for different laser output power levels. Due to the characteristics of RRM2 (lower resonator Q-factor) and the reduced rise and fall times of the chip scale driver, the signal Q-factor turned out to be less sensitive to optical carrier detuning than in the experiments reported in the previous section in a range common to both data rates, so that a common optical carrier detuning of 7.5 GHz was used for both data rates, leading to the RRM characteristics already described at the end of section II.A.2: an extinction of 8.9 dB and a reduction of the OMA by 1.4 dB from the maximum achievable (-7 dB for RRM2). ISI penalties were extracted from the recorded eye diagrams to be respectively 0.3 and 1 dB at 14 and 25 Gbps.

Based on Eq. (1)-(5) the measured Q-factors were reconstructed as in the previous subsection. The excellent match between measured (points) and calculated (lines) Q-factors shown in Fig. 14(a) further confirms our model.

In a second step the dependence of the BER on $Q_{sig}$ is compared to the theoretical expectations based on the simplified assumption of Gaussian noise statistics (black line in Fig. 14(b))

$$BER = \frac{1}{2} erfc\left(\frac{Q_{sig}}{\sqrt{2}}\right) \quad (6)$$

where *erfc* is the complementary error function.

The BER was recorded for an optimized sampling time and threshold (matching the assumptions used for the Q-factor extraction). However, as we can see in Fig. 14(b) there are significant discrepancies at both 14 and 25 Gbps between the measured BER and the BER predicted based on the measured $Q_{sig}$. While the BER at 25 Gbps is worse than expected, it is, surprisingly, better at 14 Gbps. While it is a well-known fact that the assumption of Gaussian noise statistics is not fully accurate for ASE noise and leads in particular to the wrong prediction in regards to the optimum decision threshold [40], the discrepancy in regards to the predicted BER is very slight and not sufficient to explain what we observe here [40], [41]. In order to verify this, we recorded the BER for different decision thresholds for a laser power achieving a BER in the range of 1e-12 at the optimum threshold (blue curve in Fig. 15(a)). The recorded data was then fitted assuming the non-Gaussian Chi square noise statistics derived in [40] and the extinction recorded from the experiment (red curve). The black curve in Fig. 15(a) shows the BER as a function of decision threshold assuming Gaussian noise statistics with the same 0- and 1-level ASE noise std. As can be seen the model based on non-Gaussian statistics is in close agreement with the recorded data while the optimum threshold predicted by the Gaussian noise model is significantly off. Nevertheless, the resulting BER is nearly the same for both models. The excess noise of the oscilloscope was also independently measured and found to be negligible in these experiments.

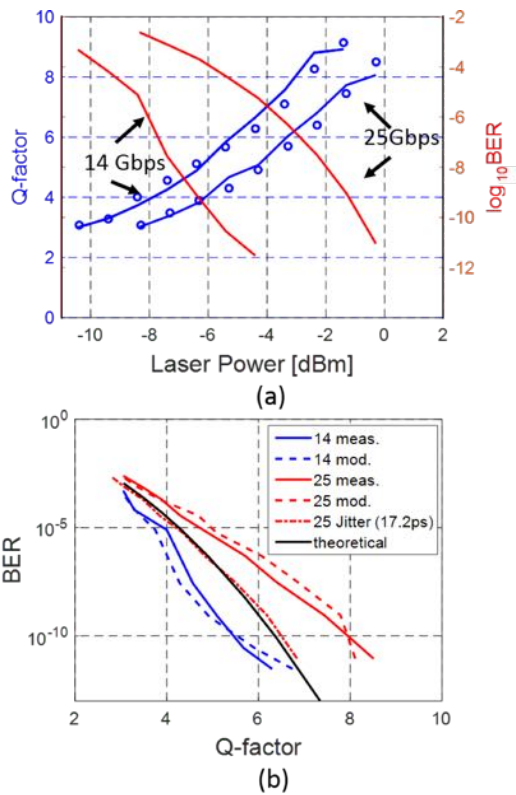

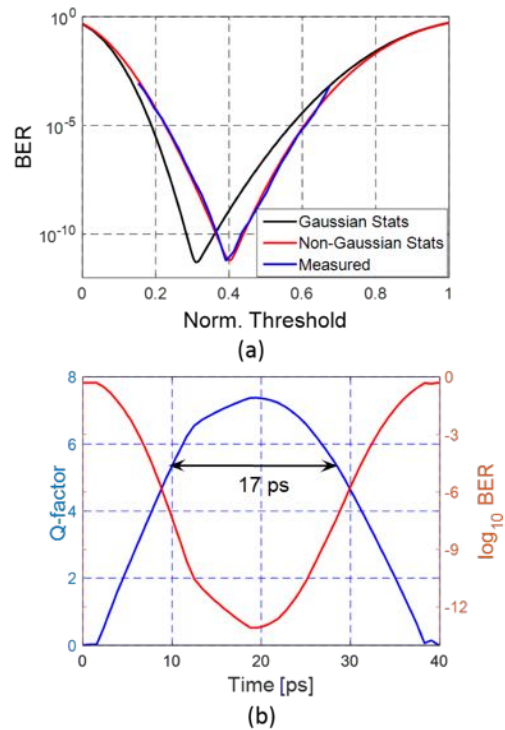

Fig. 14. (a) Measured (blue points) and modeled (blue lines) signal Q-factors and measured BER (red lines) vs. Laser Power at 14 and 25 Gbps. (b) Dependence of the measured BER on signal Q-factor for both data rates and comparison with the theoretical curve derived from Gaussian noise statistics (black line). The continuous blue and red curves correspond to the measured BER vs. the Q-factors measured with the real time oscilloscope. The measured BER is also plotted against Q-factors modeled assuming the actual ED bandwidth (dashed curves). The dash dotted red curve shows the modeled Q-factor at 25 Gbps assuming in addition a system jitter of 17.2 ps further discussed in the main text.

Fig. 15. (a) Vertical Bathtub Curve: BER measured at 14 Gbps as a function of decision threshold (blue curve), model assuming the non-Gaussian noise statistics from [40] (red curve) and model assuming Gaussian noise statistics with the same 0- and 1-level ASE noise std (black curve). The normalized decision threshold is indicated relative to the 0- and 1-levels, i.e., a threshold of 0 corresponds to the 0-level and a threshold of 1 to the 1-level. The extinction in this experiment was -7 dB. (b) Horizontal Bathtub Curve showing the signal Q-factor as well as the predicted BER, both extracted from a typical eye diagram at 25 Gbps, as a function of sampling time.

As already discussed above, one might also expect part of this discrepancy to be due to the fact that the analog front end of the BERT has a significantly higher bandwidth than the oscilloscope (32 GHz vs. 20 GHz) in combination with ASE being the primary source of noise. Indeed, ASE-signal beat noise is effectively filtered by a 20 GHz filter in either case (since half the passband of the optical filter interposed between the SOA and the Rx is 20 GHz). However, level dependent ASE noise is further modified by the 20 GHz electrical domain filtering occurring in the real time oscilloscope during the recording of the eye diagrams, further reducing the signal quality as evidenced in Fig. 10, while this effect is minimal in the case of the BERT due to its increased analog bandwidth (see Table I) on the order of half the optical filter passband (20 GHz) added to the maximum signal frequency (approximated as twice the Nyquist frequency, i.e., 14 GHz). The dashed blue curve shown in Fig. 14(b) shows the measured BER vs. the modeled Q-factor with this modification to the level dependent ASE noise taken into account. While the corrected Q-factor is then slightly better than when it is modeled with the coefficients from Table I corresponding to the 20 GHz oscilloscope filter (not shown), the difference is also too small to account for the better than expected measured BER at 14 Gbps – this is one aspect of the system that we could not fully ellucidate even though it proved to be consistent over repeated measurements.

On the other hand, the unexpected degradation of the BER vs. signal Q-factor seen at 25 Gbps does not seem to be a characteristic of the investigated link, but rather related to the test equipment as it was also seen with the same magnitude in a reference experiment consisting in cascading a noisy light source (low power laser amplified by an EDFA) with a commercial MZM directly driven by the PPG and directly fed into the U2T/Finisar Rx without post-modulation optical amplification. This experiment was compared to the investigated link (RRM + Driver Chip + SOA). The laser power was independently set in both experiments so as to obtain a low speed BER of 1e-9 and in both cases the BER worsened from 1e-9 to ~1e-6 as the data rate was increased from 14 to 25 Gbps.

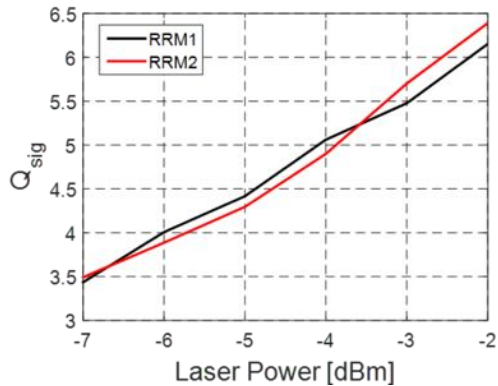

Fig. 16. Comparison of the Q-factor measured at 25 Gbps as a function of laser power for RRM1 and RRM2. RRM1 was directly probed with high-speed probe tips, while RRM2 was wire bonded to the chip scale driver.

A likely explanation for the degradation of the BER at 25 Gbps is slowly varying jitter not seen in the oscilloscope traces that are comparatively short relative to the gating periods of the BER measurements (that were done without a CDR at the ED). In order to evaluate how much jitter would be needed to explain the difference between the measured $Q_{sig}$ and the lower $Q_{sig}$ expected based on the BER measurements we extracted Q-factors from the eye diagrams that were effectively averaged over sampling time ranges according to an assumed peak-to-peak jitter: For each of the sampling times falling within this time range the corresponding BER was first calculated, averaged over the entire range and then reverse transformed into an averaged effective $Q_{sig}$. The red dot-dashed line in Fig. 14(b) shows the result when we assume a peak-to-peak time window of 17.2 ps. This amount of jitter is not unlikely since the phase margin of the BERT is specified as 28 ps corresponding to a jitter of 12 ps at 25 Gbps (the jitter of the PPG alone already accounts for up to 8 ps according to specifications). In the case of the RRM + Driver Chip system experiment this is further compounded by any jitter generated in the driver chip as well as in the commercial photoreceiver. Some additional jitter may also arise from the time delay introduced by the RRM. Since it depends on the optical carrier detuning it may vary if the setup drifts over time (due to the inaccessibility of the thermal tuner pins, an active control loop of the RRM resonance was not implemented). An exact expression of the time delay can be obtained by taking the derivative of the phase of the RRM's E/O $S_{21}$ [3] in respect to the angular RF modulation frequency. Its value at RF frequencies close to the optical carrier detuning can however be estimated as $2Q/\omega_0$, where $Q$ is the Q-factor of the resonator and $\omega_0 = 2\pi f_0$ is the angular carrier frequency, which in this case amounts to 7 ps. While the RRM cannot significantly drift from its operating point without severely compromising OMA and other RRM performance metrics, a slow drift of the time delay by 1 or a few ps is possible, further shifting the BERT sampling time from its optimum in the absence of a Rx CDR. Both experiments, the system experiment with RRM and driver chip and the reference experiment with the commercial MZM, have in common that the BER is quite sensitive on sampling time even close to its optimum and is thus sensitive to timing jitter. An example of a horizontal bathtub curve is shown for the RRM + Driver + SOA system experiment in Fig. 15(b).

The penalty observed at 25 Gbps corresponds to slightly more than 1 dB (at a BER of 1e-12 the measured Q-factor is ~9, as extracted from the eye diagrams recorded on the oscilloscope).

In order to visualize possible penalties arising from Tx integration, Fig. 16 shows a comparison of $Q_{sig}$ at 25 Gbps as a function of laser power measured with either RRM1 and instrument grade electronics connected to the SiP chip via probe tips or with RRM2 wire bonded to the driver chip. A direct comparison is made more challenging by the fact that the two modulator chips have somewhat different characteristics. While RRM1 and RRM2 were both operated with the same optical carrier detuning (7.5 GHz) for this experiment and had comparable bandwidths, RRM1 had both a better RRM OMA (-7.1 vs. -8.4 dB, resulting in a slightly higher signal level notwithstanding slightly worse cumulative GC losses of 0.6 dB) as well as a worse extinction (7.2 vs. 8.9 dB). Even though both rings have a comparable bandwidth, the extracted ISI penalty (1 dB for RRM2 vs. 0.93 dB for RRM1) is slightly higher for RRM2 pointing to a slight influence of the driver.

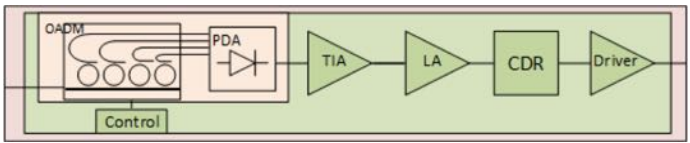

Fig. 17. Receiver architecture

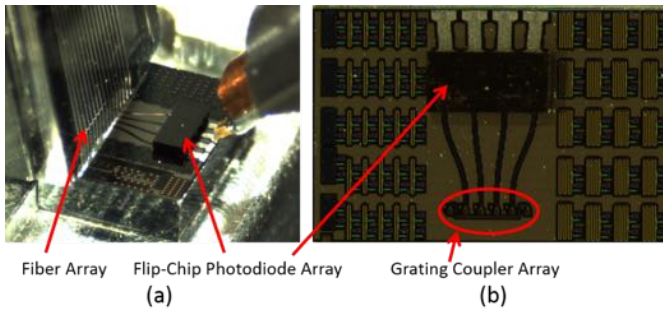

Fig. 18. (a) Optical probing of the SiP chip with a fiber array. The FC-PD is mounted on the SiP chip and electrically contacted with RF probe tips. (b) Zoomed view of the FC-PD array and SiP waveguides routed to a GC array.

Overall, one may conclude that the penalty associated to the driver chip integration has only a small influence leading to a slight increase in ISI penalty.

## III. RECEIVER

The targeted Rx architecture (Fig. 17) consists in an optical demultiplexer implemented in the form of cascaded ring based Optical Add-Drop Multiplexers (OADMs) each routing a channel to an individual drop waveguide connected to a photodiode. On-chip ILs of previously realized ring based OADMs were 0.85 dB [31]. In this architecture the OADMs also replace the second optical filter interposed between the SOA and the Rx in Fig. 8 in that they also filter out ASE and contribute to the ASE-ASE beat noise reduction.

Two versions of the Rx were implemented and characterized here, one consisting in an integrated Ge WPD directly connected to a waveguide routed to a single polarization GC acting as an optical input port. The other relies on a grating coupled FC-PD, i.e., the light first enters the SiP Rx chip via a first single polarization GC, is routed to the location of the FC-PD, and is then routed to the latter by means of a second GC. The on-chip OADMs were not yet implemented and were functionally replaced in the full link characterization (section IV) by the off-chip commercial 40 GHz notch filter with tunable center frequency also used in the previous section to select a single channel. Additional losses associated to managing polarization diversity and arising from the OADMs are separately taken into account in the link budget model described in section IV. Co-operability of both types of photodiodes with a Rx chip from Mellanox comprising a TIA, optional channel equalization (unused here), a LA and an optional CDR (used) is verified and the sensitivity floor of the Rx measured.

### A. Receiver Devices

#### 1) Flip-Chip Photodiodes

For the realization of the FC-PD based Rx we have opted for the PDCAxx-20-SC InGaAs/InP front side illuminated and front side contacted 1x4 photodiode array from Albis Optoelectronics. This component is flip-chipped onto the SiP Rx chips and the 20 µm diameter optical aperture of the photodiodes illuminated by means of a GC. It features a typical responsivity of 0.8 A/W in the C-band, a high bandwidth sufficient to support 28 Gbps serial data rates and a low capacitance for a vertical incidence photodiode of 100 fF.

The main challenge associated with the integration of this component is the alignment of the light sensitive areas of the photodiodes with the beams emitted by the GCs and minimization of resulting optical losses. Since the beams exit the GCs at a finite angle from normal incidence (16$^o$ in air) the optimum FC-PD to GC alignment also depends on the height of the bump bonds used for the flip-chip attachment. To accommodate different flip-chip attachment processes we fabricated a number of test structures (Fig. 18) with the position of the GC array offset by different amounts relative to the bump bonding pads. Furthermore, we simulated the RF properties of the assembled Rx and optimized the electrical routing of the photodiode signals to the edge of the SiP chip by tailoring the transmission lines so as to minimize electrical losses, impedance discontinuities and cross-talk. After a few iterations we have developed an attachment process yielding reproducible results based on 20 µm high SnAu bump bonds.

A photograph of an assembled SiP chip can be seen in Fig. 18. We have measured a reliable external (compound) responsivity (normalized relative to the power in the optical fiber) in the range of 0.17 to 0.21 A/W in the large majority of samples. This relatively low responsivity is expected as we are using two GCs (~6 dB compound losses in this chip design) to route the light to the FC-PD. After deembedding GC losses, we obtain a FC-PD responsivity of about 0.84 A/W, as expected. No saturation effects were observed up to the highest measured optical power of 1 mW (as launched into the fiber prior to fiber-to-chip coupling). Besides, the final bandwidth of the subassembly is typically in the range of 15 to 18 GHz when measured in a 50 Ω RF environment at a 2 V bias (see Fig. 19), which is sufficient to achieve the targeted serial data rate of 25 Gbps per channel.

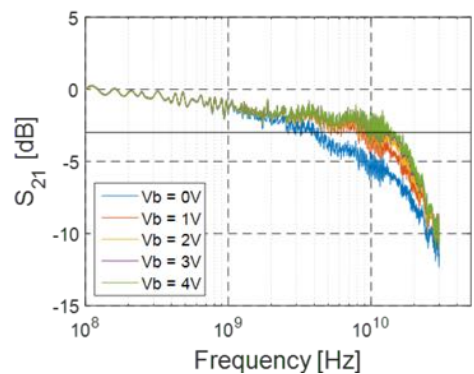

Fig. 19. E/O $S_{21}$ of the FC-PD.

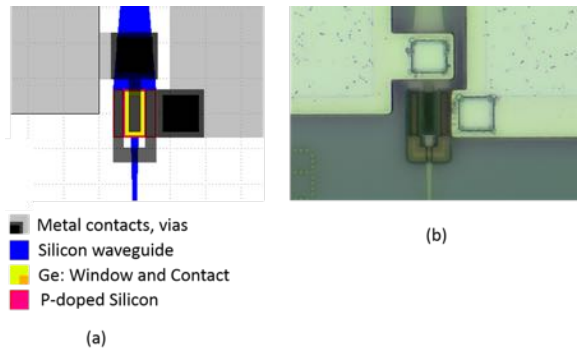

Fig. 20. (a) Layout and (b) micrograph of the Ge WPD.

*2) Germanium Waveguide Photodiode*

In parallel we evaluated SiP Rx chips with monolithically integrated Ge WPDs, with monolithic integration simplifying the assembly and reducing optical losses as well as electrical parasitics [21]. Our design is based on a reverse biased vertical P(Si)-I(Ge)-N(Ge)-junction stack (Fig. 20). The key requirements for this component are a high bandwidth, a high responsivity and a low dark current. However, these features impose some design constraints that made their mutual optimization challenging in the chosen process line at the Singapore Institute of Microelectronics (IME), A*STAR. The main challenge associated with the design of the Ge WPD was to obtain a sufficiently high cutoff frequency for 25 Gbps serial data rates, due to layout constraints associated with the metal contacts dropped onto the selectively grown Ge pads. These contact pads are required to be at least 2.4 µm to a side, constraining the sizing of the Ge pads and consequently increasing the series resistance associated to electrical transport through the underlying p-doped Si. This leads to an increased RC time constant, which turned out to be the limiting factor for the photodiodes' cutoff frequency. Reduction of this series resistance requires higher doping of the Si, which in turns results in deterioration of the material quality of the Ge overgrowth and increases dark currents. Seeking a good tradeoff between these constraints we moderately increased the p-doping of the underlying Si from the process standard to a peak implanted dopant concentration of $3e19$ cm$^{-3}$.

The Ge slab is selectively grown on the unetched Si slab and starts at an abrupt optical junction forming a discontinuity in the vertical cross-section of the waveguiding structure. The 800 nm thick Si + Ge slab is also multimode in the vertical direction, supporting a ground mode primarily confined inside the Ge as well as a first order mode with two vertical lobes. Consequently, a beating pattern arises after the abrupt interface, wherein the light periodically moves up and down between the Si and the Ge [21]. This does not only decrease the effective overlap of the light with the Ge and increase the absorption length in the device, but also allows the light to leak out of the Ge slab region when it happens to be primarily located in the Si region at its edge. We did not implement side trenches etched into the Si at the edges of the Ge stripe for additional optical confinement to avoid further increasing the series resistance to the Si contacts. This might lead to some reduction in responsivity. Furthermore, the Si-Ge heterointerface is well known to form a barrier for holes transported from the Ge to the Si region, reducing the electrical collection efficiency and increasing carrier recombination at the boundary between the two semiconductors.

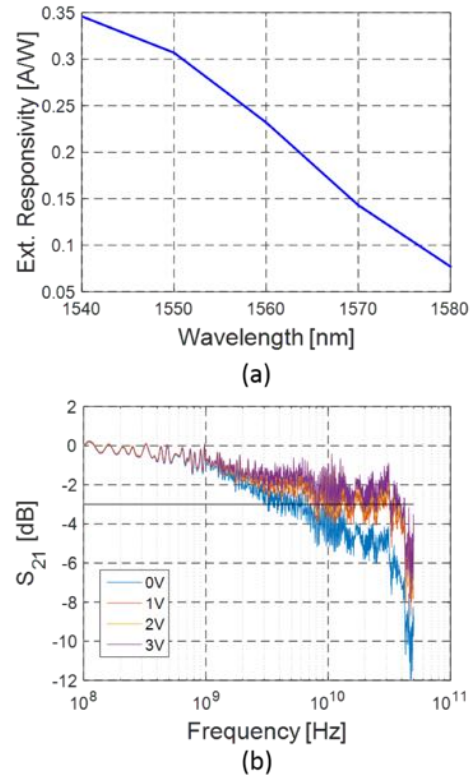

Fig. 21. (a) External (compound) responsivity (normalized relative to the power in the optical fiber) and (b) E/O $S_{21}$ of the Ge WPD probed in a 50 Ω environment.

Considering all these constraints we have designed our vertical junction Ge WPD as follows: The input waveguide is first tapered to a width of 2 µm before merging to a slab of p-doped Si (dose: $5e14$ cm$^{-2}$ at 20 keV). A stripe of low-temperature Ge with a width of 4.5 µm and a length of 12.4 µm is selectively grown in a window opened in the dielectric films over the slab-region. The top part of the Ge stripe is then n-doped using a shallow Phosphorous implantation process forming the vertical PIN-junction (dose: $4e15$ cm$^{-2}$ at 10 keV). Highly p-doped wells are defined in the Si for p-side contacting on both sides of the Ge stripe with a spacing of 200 nm from the Ge edge (dose: $4e15$ cm$^{-2}$ at 20 keV). Finally, the n-doped Ge is contacted by dropping contact plugs directly onto the stripe.

Multiple photodiodes have been measured showing a good performance with a deembedded on-chip WPD responsivity of 0.67 A/W at 1550 nm (extracted from the 0.31 A/W external responsivity shown in Fig. 21(a)) typical for vertical heterojunction Ge photodiodes [42] (even though improved responsivities have also been achieved [43], [44]). Here too, no saturation effects were observed up to 1 mW fiber power, the maximum measured optical power in the responsivity curve. The bandwidth is RC-limited and in excess of 30 GHz at 1V reverse bias when measured in a 50 Ω RF environment (see Fig. 21(b)). At 2V, the WPD capacitance is measured as 14.2 fF based on breakout structures with larger surfaces (and thus a reliably measurable capacitance). However, a typical dark current of 0.7 µA has been measured at 2 V reverse bias. From these results it is clear that the bandwidth is much higher than needed for 25 Gbps, so there is some margin to sacrifice bandwidth and improve the other performance metrics for future 25 Gbps chip iterations (in particular reduce again the dark current by reducing the doping of the underlying Si).

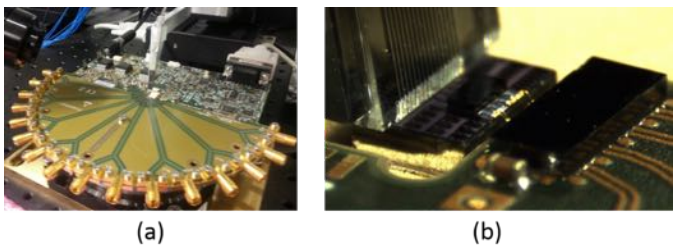

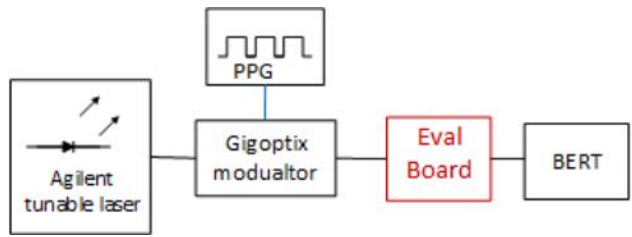

Fig. 22. (a) Photograph of the Rx evaluation board. (b) Photograph of the SiP chip with FC-PD. The SiP chip is mounted on the evaluation board and wire bonded to the TIA+LA chip.

Fig. 23. Block diagram of the measurement setup used for characterizing the Rx evaluation boards.

Concerned by the high dark current of the Ge WPDs, we characterized their noise spectrum in order to ensure that flicker noise is not an issue. We measured their noise under different illumination conditions as well as with and without voltage bias. For this purpose we contacted the photodiode chips with RF-probes and biased them to 2 V reverse bias with a bias-T. No flicker noise was discernible above the noise floor of the measurement down to the 40 kHz low frequency cutoff of the bias-T. From the noise floor of the measurement we conclude that the flicker noise power spectral density has to be below $4 \cdot 10^{-11} [\text{mW}]/f_s$ (where $f_s$ is the RF signal frequency) and the std of the total flicker noise current integrated above 40 kHz (which is below the lower cutoff frequency of the channel, 100 kHz) has to be below 0.1 µA. Given the input referred noise of the TIA (see next section), the WPD flicker noise thus does not play a significant role in the Rx sensitivity notwithstanding the high dark current of the devices and the reduced Ge material quality resulting from heteroepitaxy [45].

B. Receiver Subsystems

1) Receiver Integration and Noise Floor Measurements

In order to test the interoperability of both solutions (FC-PD and Ge-PD) with the IPTA28G4CPT Rx chip from Mellanox Technologies we mounted the SiP chips on evaluation boards and wire bonded them to the electronic Rx chip. In both cases an array of 4 photodiodes was electrically connected to the TIA + LA chip with 4 pairs of interleaved GS wires with a diameter of 25 µm bridging a distance of 200 µm between pad frames and with a distance of 130 µm between wires. The TIALA chip supports serial signaling rates up to 28 Gbd per channel with dynamically adjustable decision threshold (offset compensation), optional retiming (CDR), and transmission line drivers with configurable output swing and pre-emphasis to compensate for RF signal distortion in following PCB traces. Optional equalization of the E/O channel was not required and not used. The input impedance of the TIA is 50 Ω and the reverse bias applied to the photodiode 2 V. A photograph of the Rx evaluation board and a zoomed in view of a SiP + FC-PC + TIALA subsassembly are shown in Fig. 22.

These evaluation boards have been characterized using the setup shown in Fig. 23. An optical carrier with a 1540 nm center wavelength is generated with the tunable laser from Keysight Technologies and is coupled into a 40 Gbps MZM from Gigoptix (LX8401) with an analog cut-off frequency of 33.5 GHz serving as a reference Tx. The data stream is generated by the Anritsu PPG (MU183020A 28G/32Gbps PPG) as a PRBS-7 bit sequence fed into the modulator. The modulator is biased so as to achieve maximum extinction (> 50 dB) meaning it is not exactly biased at its 3 dB operating point but at a lower average output power to accommodate the finite drive voltage. The modulated light is directly fiber coupled to the Rx SiP chip using a GC. The electrical output of the evaluation board is connected to the 32 Gbps ED from the Anritsu BERT (MU183040A 28G/32Gbit/s ED).

Our tests (Fig. 24) show that the Ge WPD evaluation board presents better results with a sensitivity of -13 dBm at 14 Gbps (around -16.5 dBm discounting GC losses) and -9.7 dBm at 25 Gbps (-13.2 dBm discounting GC losses), defined as the average power required to achieve a BER of 1e-12 assuming infinite extinction, i.e., ½ of the required OMA. The FC-PC based SiP chips on the other hand feature a sensitivity limit of respectively -11.8 dBm and -9.1 dBm at 14 and 25 Gbps (average power in fiber). The performance enhancement of the Ge WPD based chip, 1.2 dB at 14 Gbps and 0.6 dB at 25 Gbps, is not quite as large as the expected 2 dB based on external responsivity (based on an external photodiode responsivity of 0.34 A/W and 0.21 A/W, respectively measured at 1540 nm for the Ge WPD and the FC-PD based Rx chips prior to wire bonding to the TIAs).

The discrepancy between the expected and actual improvements resulting from replacing the FC-PD with a Ge WPD might be the consequence of an additional Johnson-Nyquist noise penalty resulting from a combination of the increased Ge WPD series resistance (estimated as ~330 Ω and quite a bit larger than the FC-PD resistance estimated as 40 Ω) and the increased NEB of the Ge WPD based Rx. Indeed, the Ge WPD RC time constant results in a pole at ~30 GHz, substantially higher than the 17.5 GHz bandwidth of the FC-PD, resulting in a significantly increased aggregate NEB notwithstanding the bandwidth limitation of downstream stages (the aggregate Rx NEB is increased from ~20 to ~26 GHz which in itself already accounts for a ~0.6 dB noise penalty). As previously mentioned, the over-specified Ge WPD bandwidth opens opportunities for redesigning the Ge WPD and co-optimizing it with the TIA.

The very different sensitivities at 14 and 25 Gbps also constitute an anomaly as the system bandwidth was the same in both cases (the bandwidth of the TIALA is not adjustable). Since the output of the TIALA chip is reshaped (and the system measurement thus less sensitive to sampling jitter in the ED) and retimed (and the system measurement thus less sensitive to upstream jitter), one would expect this penalty to be of a different nature than the one seen during 25 Gbps Tx characterization. Moreover, its magnitude is much higher here (2.65 dB) so that we expect it to be primarily due to a limitation of the chip scale Rx electronics.

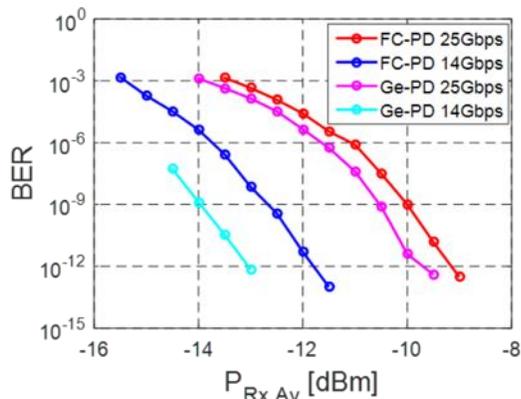

Fig. 24. BER vs. average Rx input power at 14 and 25 Gbps for the FC-PD and Ge WPD. The Gigoptix reference modulator was operated with an extinction ratio above 50 dB.

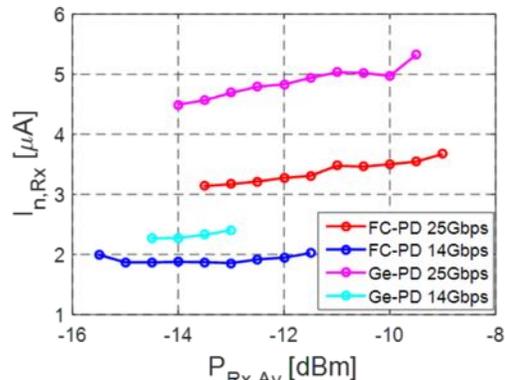

Fig. 25. Input referred Rx noise for the FC-PD and Ge WPD photoreceivers at 14 and 25 Gbps.

*2) Input Referred TIA Noise and Data Rate Dependent Penalty*

The BER data reported in the previous section can be used to extract an equivalent input referred Rx noise. The BER is first converted into a Q-factor according to Eq. (6) which is combined with (1) to extract the std of the Rx noise. Since the Rx noise is additive we assume $\sigma_1$ and $\sigma_0$ to be equal. $P_1-P_0$ is the absolute OMA (twice the average power in this case as we applied a signal with a very high extinction). We thus obtain the following expression for the std of the Rx noise

$$\sigma_{Rx} = \frac{\eta(P_1 - P_0)}{2\sqrt{2} \cdot erfc^{-1}(2BER)} \quad (7)$$

which is multiplied by the GC insertion efficiency and the photodiode responsivity in order to obtain the input referred TIA current noise (integrated over the entire equivalent noise bandwidth of the system), $I_{n,Rx}$. As a simplifying assumption in the extraction of the input referred noise, we assume there to be no additional ISI penalty arising inside the TIA, since we cannot extract it from the reshaped signal at the output of the LA (we expect this to be an adequate assumption at 14 Gbps at which the ISI penalty should be very small). The results are depicted in Fig. 25 for both the FC-PD and Ge WPD at 14 and 25 Gbps. The values at 14 Gbps can be taken to correspond to actual input referred noise integrated over the NEB, while the values at 25 Gbps also effectively take the penalty arising from data rate dependent signal distortion inside the Rx into account. This is also reflected by the fact that the extracted input referred noise levels at 14 Gbps are independent of optical power levels, as expected from additive noise, while they depend on power at 25 Gbps, pointing to the fact that they are effective values also comprising other penalties arising from e.g. bandwidth limited signal distortion or internally generated jitter (a proportional vertical eye height reduction results in a penalty that scales with the signal level on this graph – this corresponds to the 2.65 dB penalty reported in the previous section).

## IV. FULL LINK

After separate analysis of the Tx and Rx, this section is dedicated to the characterization and modeling of the whole link. The results are compared with predictions based on the Tx and Rx characteristics determined above. Due to the lack of access to the thermal tuners of the RRMs, the experimental characterization in section IV.A also relies on the bench top tunable laser. The link penalty associated to the RIN of the MLL (as experimentally investigated in section II.B.1) is reintroduced in the full link model derived in section IV.B.2. Data distortion due to SOA saturation is experimentally investigated in section IV.B.1 and the corresponding data used in IV.B.2 to assess the maximum channel count supported per SOA. Inter-channel cross-talk due to Four Wave Mixing (FWM) inside the SOA and spectral overlap between the channels is further assessed and introduced into the link budget.

### A. Combined Testing of Transmitter-SOA-Receiver Link

Figure 26 gives an overview of the test setup used for characterizing the complete link. We are characterizing the combination of Tx, SOA and Rx with a fiber coupled tunable laser used as a light source. As a consequence, we do not need to tune the RRM. Tx chip 2 (RRM2, wire bonded to the driver chip) and the Rx version with FC-PD (wire bonded to the TIALA chip) are used here. ASE-signal beat noise is filtered by the 40 GHz optical filter and the 20 GHz NEB of the FC-PD + TIALA chip (the noise mixing coefficients reported in Table I thus also apply here). With the setup seen in Fig. 27(a) we recorded BERs for laser power levels between -10 and 1 dBm at both 14 and 25 Gbps. For both data rates we biased RRM2 at a very similar optical carrier detuning than in section II.B.2 (8.9 dB extinction) so that the RRM characteristics reported there can also be assumed in this section.

The recorded BERs can be seen as dots in Fig. 27(b) together with a prediction (continuous line) based on the characterization results of the previous sections and modeling based on Eqs. (1)-(6). In the link model we assume the experimentally determined component characteristics (interface losses, RRM extinction and OMA, SOA gain and NF, FC-PD responsivity) as summarized in Table II. The input referred noise of the TIA corresponds to the data extracted at 14 Gbps from the FC-PD Rx characterization (deep blue curve in Fig. 25) and corresponds to an estimated input referred noise density of 14 $pA/\sqrt{Hz}$. An additional ISI penalty of 2.65 dB is assumed at 25 Gbps, which corresponds to the excess penalty seen at 25 Gbps in the sensitivity floor of the FC-PD Rx (red 25 Gbps curve in Fig. 24 relative to the deep blue 14 Gbps curve). Based on these numbers, the Q-factors are calculated first and subsequently converted into a BER assuming a Gaussian noise model (Eq. (6)).

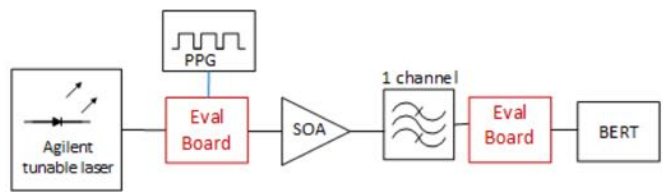

Fig. 26. Test setup used to characterize the Tx and Rx operated together.

TABLE II
LIST OF DEVICE AND SYSTEM CHARACTERISTICS (FULL LINK)

| Quantity | Value | Comment |
| --- | --- | --- |
| GC IL (Tx) | 4.8 dB | Per GC |
| RRM Ext. | 8.9 dB | |
| RRM OMA | -8.4 dB | As defined in II.A.2 |
| SOA Gain | 24 dB | SOA + up & downstream isolators |
| SOA NF | 10 dBe | SOA + up & downstream isolators |
| 40 GHz Filter IL | 1.6 dB | |
| FC-PD Ext. Resp. | 0.21 A/W | Includes 6 dB excess Rx GC losses |
| TIA Inp. Ref. Noise | 2 µA | Integrated over Rx bandwidth |
| ISI Penalty (25G) | 2.65 dB | Rx signal distortion @ 25 Gbps |

As can be seen in Fig. 27(b), the model reproduces the experimental data very well. Moreover, this data shows that no additional penalty arises from jointly operating the SiP Tx with the SiP Rx, i.e., the performance of the Rx in terms of BER as a function of data rate and input optical signal Q-factor is the same as recorded with the MZM based reference Tx. Importantly, the 2.65 dB penalty experimentally determined in section III.B for the 25 Gbps data rate fully accounts for data rate dependent penalties and is not cumulative with the data rate dependent penalty determined in section II.B.2 during Tx characterization. This is consistent with the hypothesis of the latter being related to system jitter: As the Rx chip contains a CDR and its electrical output is reshaped it is much less sensitive to PPG/Tx and ED jitter (the former being compensated by the Rx CDR and the latter having much less impact on the sampling of a reshaped data stream with flat-top symbol shapes).

### B. Complete Link Budget

In this section we construct a complete link budget and extrapolate the capability of the technology based on the experimental data reported in the previous sections. The objective is twofold:

On the one hand we have not been able to measure the complete link while both using an MLL as a light source and integrating the electronics, due to the unavailability of a pin out for the thermal tuners in the current revision of the test boards. As a consequence, the noise penalty associated with RIN has to be reintroduced into the full link model. Moreover, penalties related to multi-channel operation such as channel cross-talk and SOA saturation have not been investigated experimentally for the aforementioned reason. The associated penalties are also introduced into the full link model in section IV.B.2 after

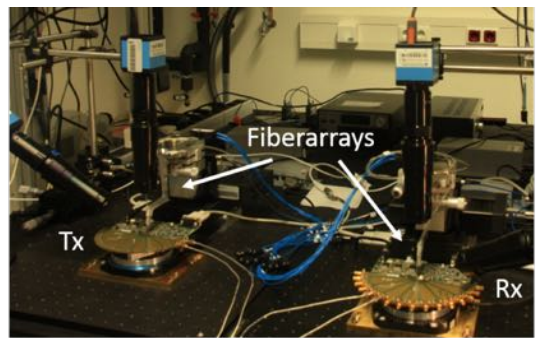
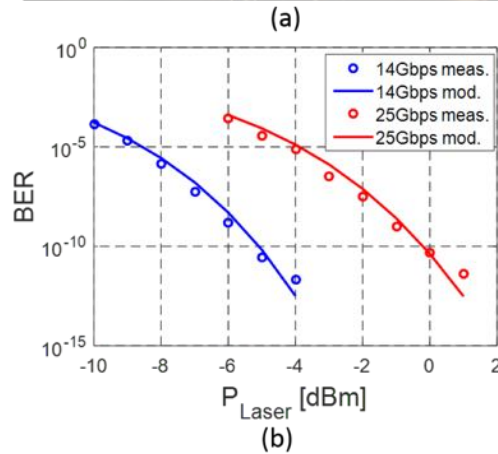

Fig. 27. (a) Photograph of the test setup combining the Tx and Rx evaluation boards and (b) measured BER as a function of laser power (points) as well as predicted BER based on the link model (continuous lines).

independent characterization of SOA saturation and FWM in section IV.B.1.

The second objective is to determine the required MLL performance and the maximum laser to SiP chip and SOA to SiP chip coupling losses compatible with the targeted channel count and data rate, as targets for further development.

It should also be noted that impairments due to fiber dispersion are not analyzed here, as the focus of this work lies on short distance Datacom links. A comprehensive study of RRM induced chirp and related limitations on long distance transmission can be found in [46].

#### 1) Effect of SOA Saturation and Four Wave Mixing

Depending on the power entering the SOA and the amount of light the SOA can generate (its output saturation power), the SOA operating point might leave the linear regime. Unlike EDFAs, SOAs have a much shorter carrier lifetime, below a nanosecond, so that the gain relaxation is sufficiently fast to track the data stream when the SOA enters saturation. This leads to a high pass behavior with a higher output modulation amplitude at signal frequencies above a transition frequency in the GHz range. In order to quantify this effect, we measure the $S_{21}$ of the utilized packaged quantum well SOA by amplifying a signal (8 dB extinction) generated with a tunable laser source and the commercial MZM modulator. Before entering the commercial photoreceiver the signal is attenuated by 16.5 dB to avoid saturation of the Rx. The average power entering the packaged SOA is varied from -19.8 to 2.2 dBm to record different levels of saturation. The data is normalized by measuring the $S_{21}$ of the same link after removing the SOA (with the optical attenuation adjusted to ensure the same average power enters the photoreceiver, since we found the cutoff frequency of the commercial Rx to be input power dependent). The measurements are fitted with solutions to the following differential equation by adjusting the gain relaxation time constant $\tau_c$

$$\frac{dG}{dt} = -\frac{G-G_0}{\tau_c} - \frac{GP_{SOA,in}}{P_{sat,in}\tau_c} \qquad (8)$$

where $G_0$ is the small signal gain at low average input powers, and $P_{sat,in}$ is the saturation input power of the SOA at the 3 dB gain compression point. $P_{sat,in}$ is -6 dBm for the packaged SOA (-9 dBm for the bare die accounting for fiber coupling losses and an interposed isolator). A $\tau_c$ of 0.17 ns was found to result in a good agreement with the measured $S_{21}$ curves displayed in Fig. 28.

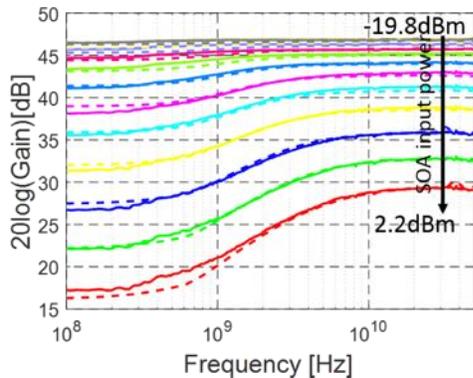

Fig. 28. Measured (continuous lines) and modeled (dashed lines) $S_{21}$ of the SOA operated with a single channel for different average power levels reported at the input of the package SOA (8 dB signal extinction).

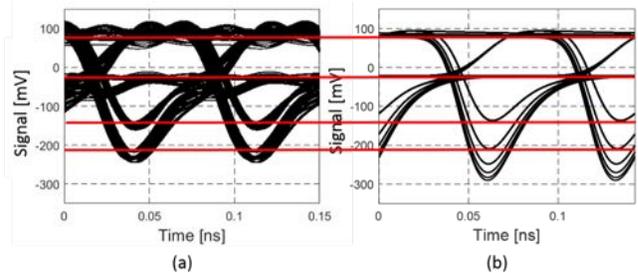

Fig. 29. (a) Measured and (b) calculated eye diagram after single channel amplification by the SOA for an average power of 1.7 dBm and a signal extinction of 18 dB, both specified at the input of the packaged SOA.

Subsequently we measured eye diagrams at 14 Gbps with the same experimental setup (only the signal extinction was changed to 18 dB) to confirm the validity of our modeling in the time domain. The eye diagram is reconstructed with the help of Eq. (8) and converted to the electrical domain by taking the power dependent (negative) conversion gain of the photoreceiver into account. Figure 29 shows a comparison of the measured (a) and the modeled (b) eye diagrams. Generally, there appears to be a good agreement between the actual and calculated signal levels.

Determining the impact of SOA saturation on the link is not entirely straightforward since the saturation seen by one channel depends on all the channels sent through the SOA. A first approximation consists in the simplified SOA model given by Eq. (8) attributing the gain dynamics to a single time constant and further making the assumption that the gain coefficients of all the channels follow the same time dependency derived from the aggregate power entering the SOA. Physically, this is an adequate assumption as long as gain dynamics are primarily determined by Carrier Density Pulsation (CDP). It does however neglect channel specific gain variations resulting from Spectral Hole Burning (SHB) and Carrier Heating (CH), which is an appropriate assumption here since, at the time scales of the data stream, gain saturation and Cross Gain Modulation (XGM) are dominated by CDP.

Another important assumption resides in the degree of correlation assumed amongst the channels: Indeed, the worst case for XGM corresponds to all the channels synchronously transmitting an identical data stream. This is however a very rare occurrence. The vertical eye opening in Fig. 29 is determined by series of at least two consecutive equal bits: The second of two consecutive 0-bits sees the maximum amount of power level dependent gain as the signal level remained low for a sufficiently prolonged duration for the gain to relax to a high level. Similarly, the second of two consecutive 1-bits sees the minimum level of power dependent gain. The likelihood for either of these situations to occur jointly for N channels is $1/2^{2N-1}$, e.g., 1/32768 for an 8 channel configuration. A sophisticated analysis would take into account the binomial distribution describing all possible combinations over the characteristic time constant of the SOA gain relaxation. This analysis would however presuppose knowledge on the degree of independence between the channels. Here, we analyze two extremes: The worst case situation corresponding to N perfectly correlated channels, as well as the best case situation corresponding in the N channels being exactly balanced in terms of the number of transported 0s and 1s. A maximum disparity between the number of 0s and 1s synchronously transmitted amongst the channels could be guaranteed with some encoding overhead, for example by transporting 8 logical channels over 10 optical channels after applying 8B/10B encoding.

A very important benefit of keeping the overall power entering the SOA constant that is even more important than the reduction of the aforementioned vertical eye opening penalty consists in maintaining a high signal extinction independently of the degree of SOA saturation, which is obtained in the balanced case since the gain remains constant. Indeed, while the reduction of the vertical eye opening is the relevant metric to assess the reduction of signal quality due to additive Rx noise, this additional eye closure does not result in a one-to-one reduction of the signal quality in a RIN and ASE limited link: Provided we assume the NF of the SOA does not change significantly at the onset of saturation, the quality of a signal with infinite extinction would remain the same, as RIN and ASE are amplified by the same reduced gain as the 1-level signal. A reduced extinction on the other hand directly results in an increased 0-level noise, and is thus the right metric to assess the effect of SOA saturation on RIN and ASE.

The two scenarios are modelled as follows: In case of balanced channels the constant gain of the SOA is simply given by

$$G = \frac{G_0}{1+\frac{N(P_{SOA,0}+P_{SOA,1})}{2P_{sat,in}}} \quad (9)$$

Assuming an unchanged NF, RIN, ASE and signal are all reduced by the same multiplier at the entrance of the Rx. Since the Rx input referred noise remains unchanged, it is effectively increased by a factor $1 + N(P_{SOA,0} + P_{SOA,1})/2P_{sat,in}$ relative to the signal strength.

In the worst case situation, we assume the gain to be largely determined by the other *N*-1 channels and to be largely uncorrelated to the data stream of the investigated channel. The gain then drifts between two extremes given by

$$G_{min} = \frac{G_0}{1+\frac{NP_{SOA,1}}{P_{sat,in}}}, G_{max} = \frac{G_0}{1+\frac{NP_{SOA,0}}{P_{sat,in}}} \quad (10)$$

Since the ASE saturation is now largely independent of the bit sequence of the investigated channel, associated penalties can be simply stacked with ISI penalties associated to bandwidth limitations in the single channel model, without having to take data sequence dependent correlations into account. The SOA saturation induced eye opening penalty (in dB) is then given by

$$-10log_{10}\left(\frac{G_{min}P_{SOA,1}-G_{max}P_{SOA,0}}{G_0(P_{SOA,1}-P_{SOA,0})}\right) \quad (11)$$

increasing, as previously, the Rx noise in relative terms (by a larger amount than in the balanced case). The reduced extinction (also in dB) is further given by

$$10log_{10}\left(\frac{G_{min}P_{SOA,1}}{G_{max}P_{SOA,0}}\right) \quad (12)$$

and now enters the link budget in the evaluation of the increased 0-level RIN and ASE noise. In the case of ASE, the reduced extinction cannot be simply referred to the SOA input and directly plugged into Eq. (4) (as is done for the case of the reduced extinction due to downstream optical filtering, as discussed in section II.B.1) since here the variable gain impacts both the signal and the amount of generated ASE. Rather, Eq. (4) is evaluated assuming different amounts of gain for 0- and 1- levels (the interaction of electrical filtering and data dependent ASE levels is not further considered here, leading to this simplification). Since the link investigated here is strongly penalized by RIN and ASE noise, the reduction of extinction turned out to be a critical factor in the link budget analysis reported in the next section.

Figure 30 shows the vertical eye opening and extinction penalties computed for both scenarios as a function of channel count (2 to 16 channels in steps of 2, assuming all other comb lines to be fully filtered out prior to reaching the SOA) and of the average power per line after coupling to the SOA chip. It forms the basis for the inclusion of SOA saturation based penalties in the next section. We also plot the data for two assumptions of pre-SOA extinction, 9 and 17 dB. When reading the graph, one should keep in mind that the typical average power per line at the input of the SOA (after coupling to the SOA chip) is below -21 dBm, assuming 0 dBm per comb line, MLL and SOA interface losses of at least 3.5 dB each, and an attenuation of the 1-level power by 11 dB by the RRM (these assumptions are discussed in the next section).

Channel penalties associated to FWM inside the SOA are also included in the link model described in the next section. Figure 31 shows the SOA characterization data on which they are based. Two lines carrying each -15 dBm of power are fed into the packaged SOA (corresponding to ~ -18 dBm after coupling to the SOA chip). This power level corresponds to the maximum expected 1-level power entering the SOA for a given channel, as discussed in the previous paragraph. Figure 31(a) shows the generated power in one of the two spectrally nearest generated lines as a function of the input channel spacing. The expected drop in FWM conversion efficiency is seen as the input channel spacing is increased from 100 to 200 and 300 GHz (the relevant spacing for the FWM induced cross-talk for the central channels in a 12 channel configuration). Importantly, the FWM efficiency also drops as a spectrally broad background (generated by the ASE of an EDFA devoid of input signal) emulating the gain saturation induced by the other channels not directly involved in a given FWM process is added to the input of the SOA. In a 12 channel configuration, this background would be larger than -8 dBm at the input of the packaged SOA assuming an average power per channel of -21 dBm entering the SOA chip (accounting for the other 10 channels as well as for the 3 dB coupling losses at the input of the packaged SOA) and is even higher in practice due to the partial extinction of a channel selection filter interposed between the MLL and the RRMs (as discussed in the next section) that lets some of the power of the nominally unused comb lines pass through, so that we estimate a background power of -5 dBm to be typical (-8 dBm on-chip after accounting for the fiber to SOA coupling losses, i.e., just slightly above $P_{sat,in}$).

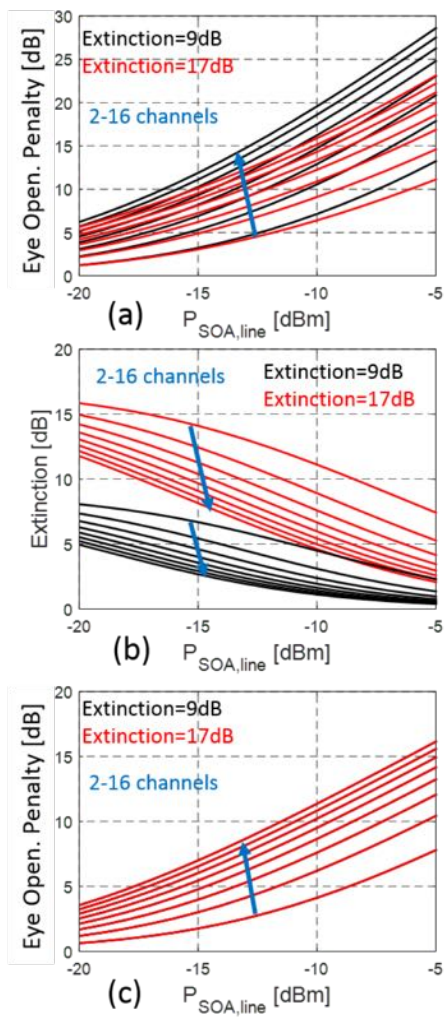

Fig. 30. Eye opening penalties due to partial SOA saturation for the worst case (a) and balanced (c) scenarios as a function of the number of channels and of the average input power per channel (specified here *after* coupling to the SOA chip). (b) shows the reduction of the extinction due to partial SOA saturation for the worst case scenario (for the balanced case the extinction is not modified, so that it is not drawn here). The black and red curves respectively correspond to an initial signal extinction of 9 and 17 dB prior to entering the SOA.

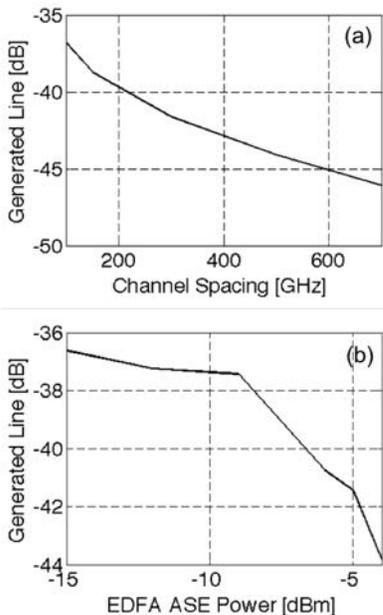

Fig. 31. Characterization of SOA FWM. (a) Power generated in each of the two spectrally nearest generated lines as a function of the frequency spacing of the two unmodulated lines each carrying -15 dBm prior to entering the packaged SOA. In (b) a wide spectrum background is added to the SOA input (ASE generated by an EDFA coupled into the signal path). The frequency spacing between the input lines is maintained at 100 GHz and the power generated in each of the nearest generated lines is plotted as a function of the background power at the input of the SOA. In both cases the power of the generated lines is indicated relative to the power of the initial lines compared to each other at the output of the SOA. FWM efficiency is seen to drop both with channel spacing and with background power. A sharp decrease in FWM efficiency is visible once a substantial background power pushes the SOA into saturation.

A worst case analysis consists in assuming the amplitudes of all the involved FWM processes aggressing a given channel to add up constructively with each other and further to interfere with the aggressed channel in such a way as to minimize the 1-level amplitude and to maximize the 0-level (these two latter assumptions are not contradictory since the RRMs also introduce phase shifts during modulation and since other elements of the system may drift over time). Deriving the beat note between the aggressed channel and the light generated from other channels by FWM, the penalty is given by

$$-10 \log_{10}\left(1 - 2\left[2\sqrt{P_{1,2}} + 2\sqrt{P_{2,4}} + \sqrt{P_{3,6}}\right] \cdot \frac{1+\sqrt{\frac{P_{SOA,0}}{P_{SOA,1}}}}{1-\frac{P_{SOA,0}}{P_{SOA,1}}}\right) \quad (13)$$

where $P_{1,2}$ is the relative power generated by two lines spaced by respectively 100 and 200 GHz from the aggressed channel, $P_{2,4}$ is the relative power generated by two lines spaced by respectively 200 and 400 GHz and $P_{3,6}$ is the relative power generated by two lines spaced by respectively 300 and 600 GHz. The amplitudes $\sqrt{P_{1,2}}$ and $\sqrt{P_{2,4}}$ are doubled to account for the fact that a pair of lines at both lower and at higher frequencies than the aggressed optical channel contribute. $\sqrt{P_{3,6}}$ is only added once, since for a central line (worst case) of a 12 channel configuration there will only be a single line spaced by 600 GHz from the optical carrier. The factor 2 before the square bracket accounts for the beating between the carrier and the FWM contributions. The last term accounts for the finite extinction of the signal at the output of the SOA (the numerator sums the reduction of the 1-level with the increase of the 0-level; the denominator normalizes the penalty to the initial OMA). Even in this worst case analysis (also taking into account a reduction of the signal extinction from 17 to 14 dB due to SOA saturation in a 12 channel configuration), the FWM penalty is estimated to remain below 0.4 dB. Overall penalties related to FWM remain low as the power contained in any given pair of comb lines is low. The high overall power entering the SOA further contributes to reducing FWM.

As a final note one should mention that reduction of RIN by means of a saturated SOA, as used in single channel configurations [47], does not occur here. This is not due to the multi-channel operation per se, but rather to the way the RIN of individual comb lines correlates with each other in a typical semiconductor comb laser: The RIN of the entire comb is orders of magnitude smaller than the RIN of isolated channels (see Fig. 3(b)), so that the SOA gain cannot be expected to correlate to any single channel RIN once a large number of channels are transmitted jointly through the SOA.

*2) Multi-Channel Transceiver Link Budget*

A number of modifications are introduced here relative to the calibrated full link model used in section IV.A to reproduce the experiments: RIN, as measured on the individual MLL comb lines and reported in section II.A.1, is reintroduced into the link model. The filter interposed between the SOA and the photodiode is assumed to be implemented in the form of a single ring based OADM integrated inside the Rx SiP chip with a Lorentzian shaped transfer function. Excess losses associated to managing polarization diversity with a Polarization Splitting Grating Coupler (PSGC) at the input to the Rx are taken into account. Penalties associated to SOA saturation and FWM as described in the previous section are further introduced into the model. An integrated wideband filter is assumed to be interposed between the MLL and the RRMs on the Tx chip in order to select optical carriers. Due to the low 1-level power transmitted by the RRM operated with 2 $V_{pp}$ (the 1-bit level is extinguished on the order of -7.5 to -11 dB by the RRM for typical optical carrier detunings), the extinction of the unused comb lines has to be quite high for them not to unnecessarily contribute to SOA saturation. A filter design based on Coupled Resonator Optical Waveguides (CROW) will be published separately. Finally, the MLL and SOA are assumed to be hybridly integrated into the Tx, with the SOA coupled back to the Tx SiP chip before light being grating coupled to the output fiber, so that it has two optical interfaces with the SiP chip. Since the MLL to SiP chip, SiP chip to SOA and SOA to SiP chip interface losses are not yet known (the integration process is still under development), these losses are parameterized in the following and the maximum acceptable interface losses derived.

We are first analyzing the effect of the OADM optical filter bandwidth on the performance of the link. A tradeoff has to be struck between ISI, a minimization of which requires a large filter bandwidth, and inter-channel cross-talk to adjacent channels, a minimization of which requires a small filter bandwidth. A smaller filter bandwidth is also beneficial in reducing the ASE-ASE beat noise. The effect on ASE-signal beat noise is less as the latter is already filtered by the O/E bandwidth of the Rx, but nonetheless here too the OADM filter bandwidth still plays a role on the overall NEB.

In this section the Tx is modeled as a driver with a finite rise and fall time (16 ps for 20% to 80% transitions as per specifications) followed by a single pole filter modelling the transfer function of the RRM. The RRM is modeled based on the measured characteristics of RRM2, but assuming a lower optical carrier detuning (4.4 GHz) than in the system experiments of the previous sections in order to obtain a larger extinction (17 dB) at the cost of a reduced cutoff frequency (18.5 GHz) and OMA (-10.9 dB). Simulations of the linear front end of the TIALA can be fitted well by a third order Butterworth filter with a 21 GHz cutoff frequency. The photodiode itself is assumed not to further limit the Rx bandwidth, as it is assumed to be the low capacitance Ge WPD described in section III.A.2.

In order to assess the inter-channel cross-talk penalty associated to the OADM bandwidth, a trapezoid shaped 25 Gbps PRBS-7 signal with the correct rise and fall times is first preconditioned by filtering it 500 times with a low pass filter with a 400 GHz cutoff frequency, yielding a smooth signal shape. It is then filtered by a single pole filter emulating the RRM. Next, the adjacent signals are generated by time shifting the PRBS signal by random amounts and up and down conversion by 100 GHz. The sum of the three signals is sent through the OADM and Rx filters after which an eye diagram is generated and its vertical opening extracted.

The OADM transfer function is modeled by a Lorentzian as

$$\frac{E_{Drop}}{E_{In}} = \frac{1}{1+i\frac{f-f_0}{\Delta f_{OF}}} \tag{14}$$

where $E_{In}$ is the amplitude of the light entering the OADM, $E_{Drop}$ is the amplitude of the light coupled to the drop port, $f$ is the frequency of the light, $f_0$ is the resonant frequency of the ring assumed to be equal to the carrier frequency of the targeted channel, and $\Delta f_{OF}$ is the equivalent single sided passband of the OADM (note that the NEB plugged into Eq. (4) is $\pi/2$ larger, since this is a single pole filter). $\Delta f_{OF}$ is half the Full Width at Half Maximum (FWHM) of the resonator and is thus given by $\Delta f_{OF} = f_0/2Q$ where $Q$ is the loaded Q-factor of the cavity. Note that the IL of the OADM are accounted for separately in the following and are thus not considered in Eq. (14).

Figure 32(a) shows the results. The computed vertical eye opening (normalized OMA, given as a fraction of the input signal) taking into account ISI arising from the 16 ps driver rise and fall time, an 18.5 GHz RRM bandwidth (single pole), the variable OADM bandwidth (single pole) and the 21 GHz Rx bandwidth (third order Butterworth), as well as taking into account channel cross-talk due to spectral overlap, is shown in red (SOA XGM, FWM, various ILs and the modulation penalty are taken into account separately in the link budget; this data only serves to determine the optimum OADM bandwidth in view of ISI and spectral channel overlap). For comparison, the vertical eye opening obtained by considering ISI only, i.e., without adding the adjacent channels, is shown in blue. The black curve shows the effect of cross-talk without considering OADM induced ISI, computed by adding the adjacent channels to the targeted channel, but applying the OADM transfer function only to the adjacent channels and not to the targeted channel. This way, the targeted channel ISI suffers only from the Tx E/O and Rx O/E filters, but not from the OADM bandwidth. The red curve combines the effect of cross-talk and OADM induced ISI and is thus limited by the blue curve (OADM ISI) for low $\Delta f_{OF}$ and by the black curve (cross-talk) for large $\Delta f_{OF}$.

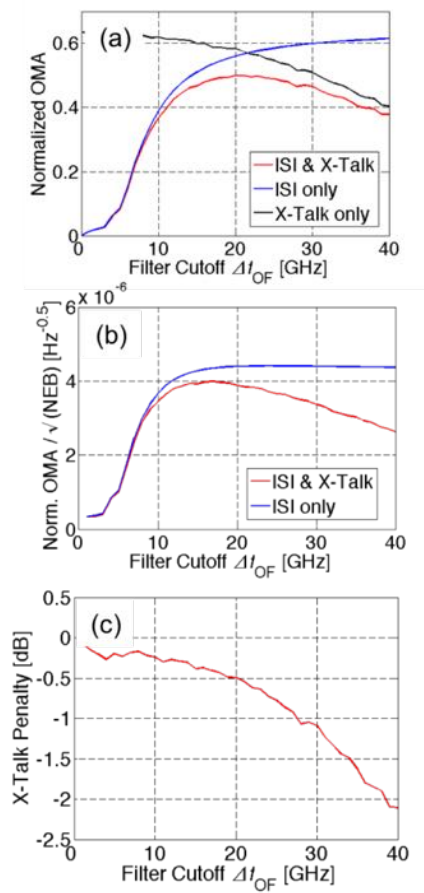

Fig. 32. Modeling of the effect of the OADM bandwidth on the channel performance. (a) shows the vertical eye height after filtering by the Tx, OADM and Rx (blue curve). The red curve also incorporates the effect of inter-channel cross-talk from adjacent channels due to the finite OADM bandwidth. The black curve corresponds to results when the targeted channel is only filtered by the Tx and Rx (i.e., ignoring OADM induced ISI) and is representative for the cross-talk penalty. In (b) the ISI only and full model curves from (a) are divided by the square root of the NEB resulting from the OADM and Rx filters. This number is characteristic for the signal quality if ASE-signal beat noise is the dominating noise term. (c) shows the cross-talk penalty at a fixed OADM bandwidth and corresponds to the ratio of the red and blue curves from (a). This curve is directly used as a vertical eye opening penalty in the full link model. In all three graphs the x-axis corresponds to the equivalent single sided cutoff frequency of the OADM ($\Delta f_{OF}$).

The best vertical eye opening is obtained at $\Delta f_{OF} = 20$ GHz (i.e., for a 40 GHz optical filter bandwidth) and is 1 dB below the highest vertical eye opening obtained with an infinite bandwidth OADM in the absence of adjacent channels. The system level penalty compared to single channel operation is however not quite as high once all factors are considered. Indeed, reducing the OADM bandwidth also contributes to reducing ASE noise. Figure 32(b) shows the vertical eye opening for ISI only and for ISI + cross-talk penalty, both divided by the square root of the NEB resulting from the OADM and the Rx filter functions. This factor would be proportional to the signal Q-factor if the noise were uniquely due to ASE-signal beat noise. The penalty between the best-case single channel operation and the operation with cross-talk to adjacent channels then reduces to 0.4 dB at $\Delta f_{OF} = 20$ GHz. Since in practice the two dominant noise sources in this link are ASE and RIN (the latter not significantly impacted by the OADM transfer function due to its comparatively low frequency nature, since the RIN decays to shot noise levels above 4 GHz), the actual penalty is in between these two values. Figure 32(c) shows the cross-talk penalty as applied to the system at a specific OADM single sided bandwidth $\Delta f_{OF}$ (and simply corresponds to the ratio of the red and blue curves in Fig. 32(a)). This function can be directly imported in the full link model and applied as a penalty to the final eye height arriving at the Rx (note that this is a bounded penalty and not a noise term). It reaches 0.5 dB at $\Delta f_{OF} = 20$ GHz. In this model, the ISI penalty right after the Tx is calculated as 1 dB at 25 Gbps, consistent with the measurements done in section II.B.2. At 25 Gbps, the ISI penalty is further 1.7 dB after the OADM and 1.8 dB after the Rx filter (to which the 2.65 dB data rate dependent Rx penalty determined in section III.B.2 is added in the link budget).

A different type of channel cross-talk resides in the weak modulation of an optical carrier by a RRM nominally assigned to a different channel due to a finite overlap of the RRM's Lorentzian transfer function at the optical carrier frequency. This inter-channel cross-talk is calculated as -25 dB between adjacent channels based on the optical carrier detuning, channel spacing and on the Q-factors of the RRMs, and is quite low due to the high attenuation applied by the main on-channel RRM. There is however a non-negligible penalty associated to the (average) attenuation of the optical carrier by the RRMs of the two adjacent channels, that cumulatively amounts to 0.5 dB additional ILs. It should be noted that channel cross-talk in cascaded RRM based systems has also been extensively experimentally studied by M. Ashkan Seyedi et al. with results reported in [48].

For managing polarization diversity at the Rx we are considering the use of a PSGC. While PSGCs have higher IL than single polarization GCs on the order of 6 dB [49] when fabricated in a standard 220 nm device layer thickness Silicon-On-Insulator (SOI) wafer, a number of strategies have been successfully implemented to reduce these ILs. Higher insertion efficiency can be obtained by modifying the thickness of the Si device layer either by modification of the SOI wafers [50] or by depositing a poly-Si layer on top of a standard 220 nm device layer [51]. Further improvements have also been obtained by increasing the directionality of the GCs with help of dual SOI stacks [50] or the deposition of a metal back-reflector [52].

TABLE III
LIST OF DEVICE AND SYSTEM CHARACTERISTICS (NEXT GEN. LINK)

| Quantity | Value | Comment |
| --- | --- | --- |
| MLL to Tx IL | $IL_A$ | Scanned |
| Tx Filter IL | 1 dB | Simulated |
| Tx Filter Extinction | 17 dB, 15 dB | Simulated for 8 & 12 channels |
| RRM Ext | 17 dB | Exp. data from Fig. 5(b) |
| RRM OMA | -10.9 dB | Exp. data from Fig. 5(b) |
| RRM Bandwidth | 18.5 GHz | Based on device models |
| IL off-channel RRMs | 0.5 dB | Calculated based on Q-factors |
| Driver Rise/Fall Time | 16 ps | From driver specifications |
| Tx to SOA IL | $IL_A$ | Scanned |
| SOA Gain @ 25°C | 27 dB | Bare die, below specification |
| SOA Gain @ 45°C | 24 dB | Bare die, 30 – 6 dB |
| SOA Gain @ 55°C | 21 dB | Bare die, 30 – 9 dB |
| SOA NF @ 25°C | 6 dBe | Bare die, as specified |
| SOA NF @ 45°C | 7.5 dBe | Bare die, 6 + 1.5 dBe |
| SOA NF @ 55°C | 8.25 dBe | Bare die, 6 + 2.25 dBe |
| SOA $P_{sat,out}$ @ 25°C | 18 dBm | Bare die, as specified |
| SOA $P_{sat,out}$ @ 25°C | 17 dBm | Bare die, 18 – 1 dBm |
| SOA $P_{sat,out}$ @ 25°C | 16.5 dBm | Bare die, 18 – 1.5 dBm |
| SOA to Tx IL | $IL_A$ | Scanned |
| FWM Penalty | 0.4 dB | Estimated from measured FWM |
| Tx to Fiber IL | 3 dB | Measured (perm. attached GC) |
| Fiber to Rx IL | 5 dB | Literature review |
| OADM IL | 0.85 dB | Measured |
| OADM FWHM | 40 GHz | Targeted ($\Delta f_{OF} = 20$ GHz) |
| Ge WPD Int. Resp. | 0.7 A/W | On chip resp., as measured |
| TIA Inp. Ref. Noise | 2 μA | From measured Rx noise floor |
| TIA Bandwidth | 21 GHz | 3rd order Butterworth, from sims. |
| Calc. ISI Penalty | 0.76, 1.8 dB | Calc. @ 14 & 25 Gbps |
| Excess Rx Penalty | 0, 2.65 dB | Measured Section III.B.2 |
| Inter-Channel X-talk Eye Opening Penalty | 0.5 dB | Calc. for 25 Gbps based on spectral overlap |

PSGC IL as low as 2.8 dB (without dual SOI stack) and 2.2 dB (with dual SOI stack), both in 310 nm device layers, have been experimentally demonstrated. Here we found that at room temperature the link budget is primarily limited by ASE and RIN due to the high gain of the SOA, so that we can assume a conservative value for the PSGC IL without excessively penalizing the link. We are assuming 5 dB IL in the following, corresponding to the 6 dB reported in [49] and a 1 dB improvement after permanent attachment with an index matched epoxy. Lower PSGC losses would of course allow dialing down the SOA gain and thus be conducive in increasing the channel count, as a same amount of SOA output saturation power could be allocated amongst a larger number of channels.

Table III summarizes the assumptions made to evaluate the link budget of the next generation of transceivers, which are currently under development. Since the SOA and laser attachment processes are still under development, a large uncertainty remains in regards to the IL that will be achievable. For this reason, no single assumption is made on the attachment losses (noted as $IL_A$ in the following). Rather, they are scanned together with MLL RIN and line power in order to map the acceptable parameter space. ASE is further modeled with the parameters from Table I. RRM characteristics assume a reduced optical carrier detuning than in the system experiments, but correspond to experimental data for RRM2.

Two items in this table require further discussion: The temperature dependent SOA characteristics and the "Tx Filter Extinction".

At room temperature, the assumed SOA gain has been *reduced* from the 30 dB specified for the commercially available bare die to 27 dB, as reducing this high gain does not penalize the channel much (at room temperature the noise budget is dominated by ASE and RIN so that Rx noise only plays a small role) while it enhances the channel count scalability given the available SOA saturation power (18 dBm output power at the 3 dB gain compression point). On the other hand, at increased temperatures we evaluate the available gain by taking the specified 30 dB room temperature as a baseline and subtracting the temperature dependent gain reduction. In other words, the available commercial SOA has adequate sizing in the increased temperature operation of a semi-cooled module. Indeed, as discussed below, at 55°C the link budget is equally limited by RIN, ASE and Rx noise given these SOA characteristics. The 6 dB gain reduction at 45°C was measured and the 9 dB gain reduction at 55°C extrapolated. The 1 dB reduction in output saturation power at 45°C is also based on a measurement, while the increase in the noise factor is based on a typical temperature dependence reported in the literature [53].

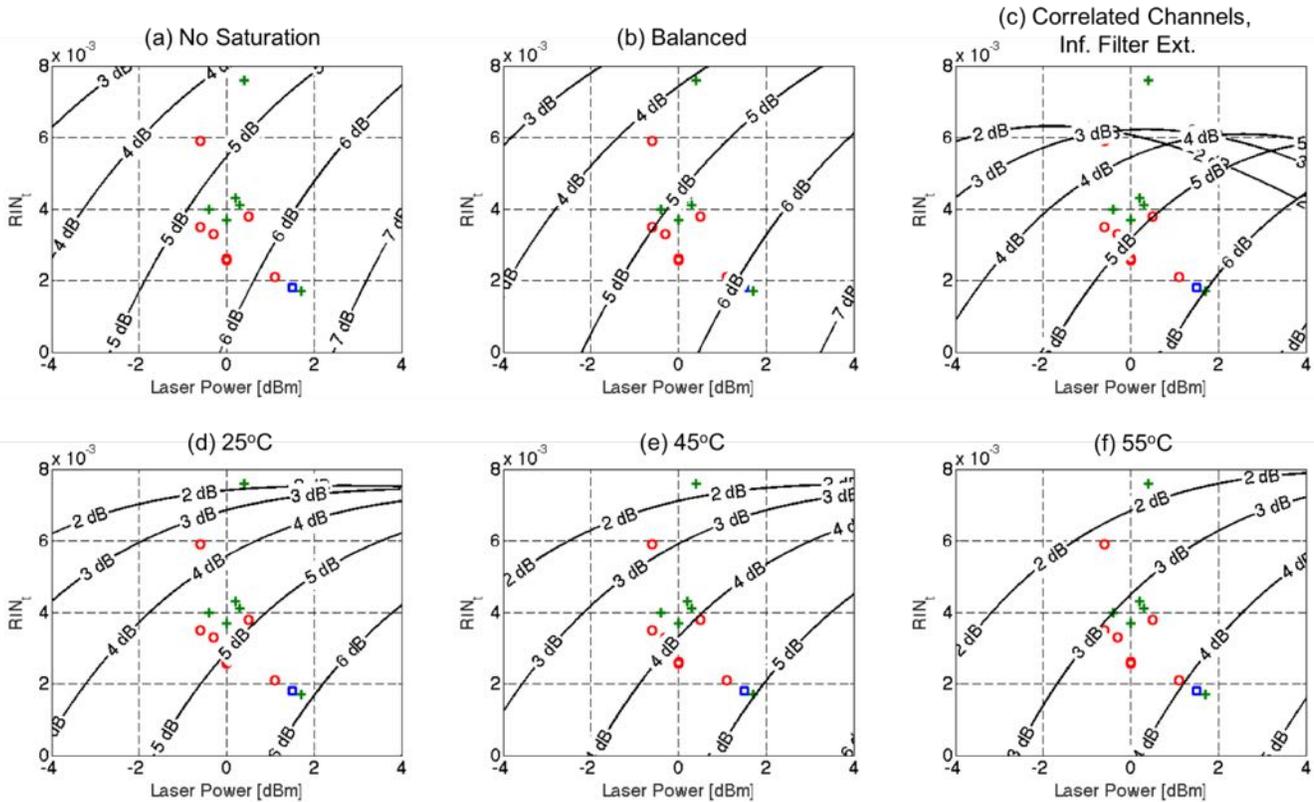

Fig. 33. Required integrated line RIN and line power as a function of the MLL to SiP and SOA to SiP interface losses ($IL_A$) at 14 Gbps serial data rates assuming (a) no SOA saturation and (b)-(f) SOA saturation with (b) 12 balanced channels and 15 dB Tx Filter extinction (c) 12 correlated channels and infinite Tx Filter extinction and (d)-(f) 12 correlated channels and 15 dB Tx Filter extinction at (d) 25°C, (e) 45°C and (f) 55°C.

The "Tx Filter Extinction" is actually a system metric that combines CROW filter and MLL characteristics: By this we refer to the amount of optical power from all the unused comb lines transmitted to the output of the filter divided by the total power of all the comb lines used as optical carriers also at the output of the filter. The finite extinction by the CROW filter of the unused lines results in an increased level of SOA saturation relative to a perfect filter. It is a metric that is a function both of the transfer function of the CROW filter as well as of the spectrum of the MLL. Its evaluation incorporates the results of a sensitivity analysis based on typical fabrication variations. It is indicated here as it enters the link budget calculation, but the details of the CROW filter design will be reported elsewhere.

Importantly, at the system level these extinctions are not as high as would first appear, as the optical carriers are attenuated by the RRMs while the unused carriers are not. If we take a 14 dB extinction of the average carrier power by the RRMs into account, these Tx Filter Extinctions (respectively evaluated for the 8 and 12 channel version of the filter) drop to 3 and 1 dB. I.e., the additional unused optical power is respectively 50% and 80% of the average "useful" optical power. Importantly, as shown in the following, this (initially unwanted) additional constant power enhances the link budget rather than penalizing it in the worst case analysis corresponding to synchronized channels, as it contributes to stabilizing the SOA input power and thus also reduces the penalization of the signal extinction.

Figure 33 shows 6 scenarios at 14 Gbps. (a) shows the maximum laser RIN and minimum required line power for different assumptions in regards to $IL_A$ assuming the SOA to be operated in the linear regime. This serves as a baseline performance of the link. The RIN and line power measurements done on the MLL are overlaid on the graph. It can be seen that $IL_A = 5$ dB, a 0 dBm line power and an integrated RIN for isolated comb lines of 4e-3 would support a 1e-12 BER, so that an implementation of the link for small channel counts appears realistic (not surprisingly given the performance of the links experimentally shown in the previous sections). The line power levels plotted in Fig. 33 contain the MLL to fiber ILs incurred during the MLL characterization (that are redundant with the MLL to chip coupling losses accounted for by $IL_A$ here). We opted not to normalize them out, since increasing the operating temperature of the laser from 25°C to 45°C would already result in a (measured) 1 dB drop of line power; this margin can thus be allocated to an increased laser operating temperature. (b) shows a 12 channel configuration assuming the balanced case, wherein the Tx Filter Extinction is assumed to be 15 dB. A number of channels could be allocated to coding overhead to reduce the bit disparity and approach a truly balanced scenario. Since the extinction is not impacted by SOA saturation in this case, only the relative weight of Rx noise is modified and the link penalties remain modest. (c) on the other hand shows the results for the worst case analysis corresponding to 12 synchronized channels assuming a perfect (infinite extinction) Tx filter. The effect of SOA saturation is now clearly visible in the inflection of the contour plots at high line power levels. Interestingly, once the finite 15 dB extinction of the Tx filter is factored back in (d), the overall situation improves again as a consequence of the stabilization of the SOA gain, i.e., the inflection of the curves occurs at a notably higher laser power.

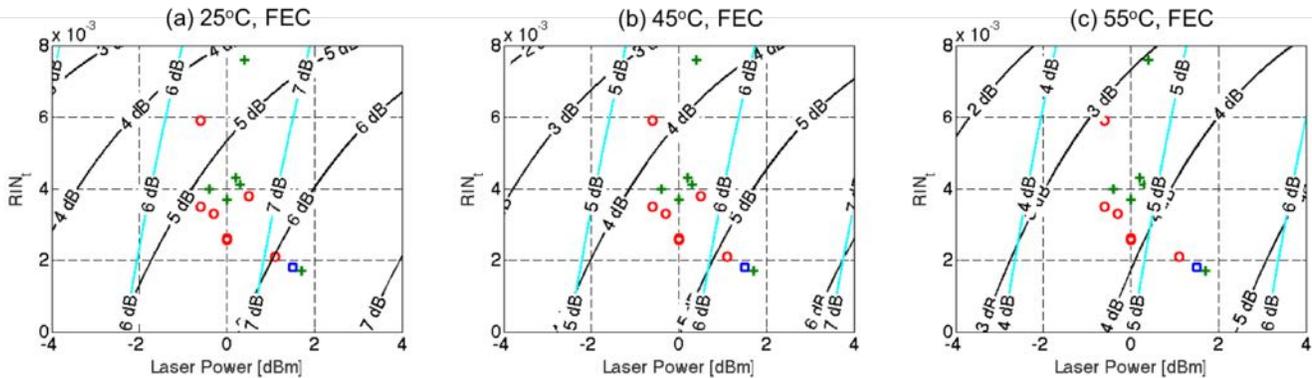

Fig. 34. Required integrated line RIN and line power as a function of the MLL to SiP and SOA to SiP interface losses (IL$_A$) for a 12 channel link with 25 Gbps serial data rates at 25°C, 45°C and 55°C assuming correlated channels (worst case) and 15 dB Tx filter extinction. The black contours show the required characteristics for 5e-5 uncorrected BER compatible with 100G Ethernet standards and the blue contours the required characteristics for 3.6e-3 uncorrected BER compatible with the 7% overhead hard decision FEC limit.

The graphs (d)-(f) show how the performance requirements for maintaining an error free link (BER < 1e-12) evolve at increased temperatures. These also correspond to a 15 dB Tx Filter Extinction as well as 12 correlated channels. The reduced gain, reduced saturation power and increased NF are factored in. At 55°C the requirements on the interface losses become very challenging in the absence of other improvements (discussed below). However, assuming interface losses of IL$_A$ = 5 dB in that scenario would still result in a signal Q-factor of 3.6 and an uncorrected BER of 1.6e-4 that is sufficiently high, with a healthy margin, to support BER < 1e-12 with standard 7% hard decision Forward Error Correction (FEC) [54]. Interestingly, at 55°C, at the typical laser RIN and line power of 4e-3 and 0 dBm, and at the required ~3.5 dB interface losses, the link is equally limited by RIN, ASE and Rx noise, with noise levels $\sigma_0 + \sigma_1$ at the input of the TIA of respectively 4.3, 4.1 and 4.0 µA (the calculated electrical modulation amplitude at the input of the TIA is 46 µA). At increased temperatures Rx noise becomes equally important as RIN and ASE due to the drastically reduced SOA gain.

Not surprisingly, the realization of a 25 Gbps WDM link is significantly more challenging: A link budget analysis shows that the current version of the chip scale electronics, combined with the SOA and MLL performance, is not sufficient for reliable 25 Gbps multi-channel operation, particularly once the 2.65 dB Rx penalty at 25 Gbps is taken into account. Under the assumption of improved electronics consisting in a reduction of the Tx rise and fall time to 8 ps and in a reduction of the 2.65 dB Rx penalty to 0.5 dB, the situation looks much improved, even though some amount of FEC still appears to be required to sustain the link: Figure 34 shows the performance requirements for a 12 channel 25 Gbps link (15 dB Tx Filter Extinction, correlated data streams) under the assumption of improved electronics (as defined above) as a function of SOA temperature in order to achieve a signal Q-factor of 3.9 and a BER of 5e-5 compatible with the Reed-Solomon FEC used in 100G Ethernet standards[3] (black contours). The light blue contours show the performance metrics required to reach a BER of 3.6e-3 that remains in principle compatible with 7% overhead hard decision FEC [55].

A number of measures can be implemented to further improve the link. Most straightforwardly, one may increase the drive voltage above 2 V$_{pp}$ or combine a dual output driver with a ring assisted MZI to reduce the modulation penalty and reduce chirp [56]. Implementing MLLs with integrated DBRs with a reduced reflection band might also allow increasing the power per line at the cost of a reduced channel count.

## V. CONCLUSIONS

We have experimentally investigated a SiP Tx/Rx pair intended to be operated in WDM configuration with a single section Q-dash MLL. Error free (BER < 1e-12) operation of the Tx/Rx pair with a single channel sourced by an instrument grade tunable laser was reached at 25 Gbps with 1 dBm of laser power. Error free operation was also reached with the SiP Tx and a single channel sourced by an MLL comb line combined with a commercial Rx. A complete multi-channel operation was not yet possible due to shortcomings of the current test fixtures. We have however reintroduced channel cross-talk, RIN and SOA based saturation into a link budget model calibrated and verified with the aforementioned single channel experiments. We conclude that 12 channel operation at 14 Gbps serial data rates is realistic in cooled operation pending MLL to SiP chip and SOA to SiP chip interface losses better than 4.5 dB (no channel pre-emphasis, equalization or FEC). Semi-cooled operation with SOA temperatures up to 55$^{o}$C appears more challenging, but still realistic with some incremental improvements. Assuming a modest improvement of the currently utilized Tx and Rx electronics, 12 channel 25 Gbps still requires FEC but is expected to be compatible with standard 7% hard decision FEC in semi-cooled operation.

---

[3]As per clause 91 of IEEE Standard 802.3bj. Based on 5 dB of coding gain, see for example the 100G-CLR4 specifications.

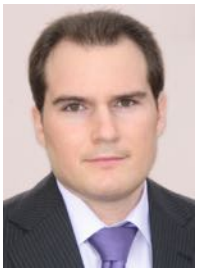
**Alvaro Moscoso-Mártir** received the M.Sc. degree in Telecommunications Engineering from Málaga University, Spain, in 2008 and the Ph.D. degree in Microwave Engineering from Málaga University in 2013. During his Ph.D. he focused on the design of ultra-wideband radiofrequency circuitry in integrated planar technology for six-port receivers. Currently, he is working as a Postdoctoral Scholar at the Institute of Integrated Photonics of the RWTH Aachen, Germany, where his main research activities focus on the design of microwave circuits and components and on the modeling and assembly of optoelectronic components**.**

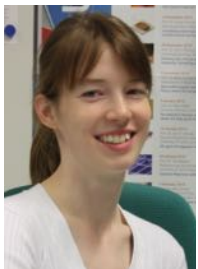
**Juliana Müller** received the M.Sc. degree in Electrical Engineering from RWTH Aachen University, Germany, in 2013 with a major in Micro- and Nanotechnologies. She is currently a doctoral candidate at the Institute of Integrated Photonics of RWTH Aachen University. Her current research focuses on Silicon Photonics WDM transceivers based on resonant ring modulators and semiconductor mode-locked lasers.

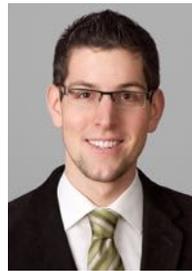
**Johannes Hauck** received the Dipl.-Phys. Degree from RWTH Aachen University, Germany, in 2011. He is currently a doctoral candidate at the Institute of Integrated Photonics of RWTH Aachen University. The focus of his current research lies with Silicon Photonics communications systems with a particular emphasis on coherent WDM transceivers. From 2005 to 2011 he studied physics at RWTH Aachen University with focus on optical systems, laser technology, extreme-ultraviolet optics and solid-state physics.

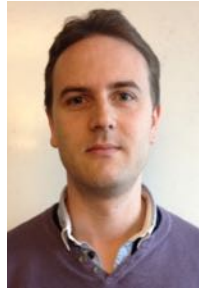
**Nicolas Chimot** received the Master's degree in Light and Matter Interaction from the University of Rennes I, France, in 2003 and the Ph.D. degree from the University of Paris XI, France, in 2006. His Ph.D. work was focused on the development of photoconductive emitters and detectors driven at 1550 nm for the generation of terahertz electromagnetic waves and the conception of ultrafast photoswitches. From 2006 to 2008, he was a Postdoctoral Researcher at the Molecular Electronic Laboratory of the Commissariat à l'Energie Atomique, Saclay, France. He worked on the conception of carbon nanotube based transistors on flexible substrates for high frequency applications and on the conception of nanoelectromechanical devices with carbone nanotubes. From 2008 to 2015, he worked at III-V Lab and Alcatel-Lucent, Marcoussis, France, as a member of the Photonic Integrated Circuit team where he was involved in the optimization and characterization of quantum dash based lasers for telecommunication applications. Since 2016 he is working at 3SPTechnologies.

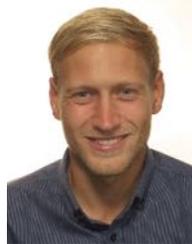
**Daniel Egede Rasmussen** received the B.Sc.E.E. from University of Southern Denmark in 2010 and M.Sc.(Eng.) in Optics & Electronics from Aarhus University, Denmark, in 2012. From 2012 to 2014 he worked as a Fellow at the European Organization for Nuclear Research (CERN) in Geneva, Switzerland. His work was on modelling of the superconducting magnets and protective circuits of the LHC in order to establish and optimize thresholds for the quench protection system. Since 2014 he has been with Mellanox Technologies, Denmark, where he is working on the development of optical transceivers for high speed communication.

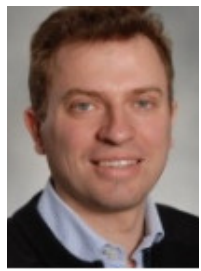
**Alexandre Garreau** received the Diploma in Condensed Material Science in 1993 from the University Paris Sud, France and the Ph.D. degree in 1997 from the University Louis Pasteur, Strasbourg, France. From 1994 to 1997, he worked for CEA at the Direction of Military Applications. His thesis subject was devoted to the fabrication and the characterization of radiation-absorbent materials. From 2000, he works as research engineer, for III-V Lab. He is focused on InP optoelectronic emitter devices, integrating several optical functions for complex modulation formats like BPSK, QPSK, QAM. He has published 30 papers and communications in national and international conferences and has written 7 patents.

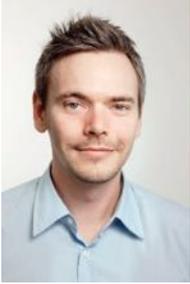

**Mads Nielsen** received the M.Sc.E. and Ph. D. degrees in electrical engineering from the Technical University of Denmark in 2000 and 2004, respectively, for work on semiconductor-based all-optical signal processing.

In 2005 he joined NEC System Devices Research Laboratories in Otsu, Japan where he was engaged in research and development of widely tunable lasers and high-speed modulators. In 2008 he joined u$^2$t Photonics in Berlin, Germany, where he worked on development of receivers for optical 40G D(Q)PSK and 100G coherent modulation formats.

Since 2013 he has been with Mellanox Technologies Denmark, currently in a position as Sr. Manager of Interconnect Design. He has published more than 70 papers in international journals and conferences.

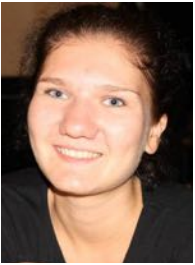

**Elmira Islamova** received the Bachelor's degree in Telecommunications from SPbNRU ITMO, St. Petersburg, Russia, in 2013 and is going to receive the M.Sc. in Communications Engineering from RWTH Aachen University, Germany, in 2016. From May to November 2015 she did her Master's thesis at the Institute of Integrated Photonics of the RWTH Aachen where she worked on the characterization of silicon ring-resonator modulators, add-drop multiplexers and semiconductor optical amplifiers. She is currently doing research of nonlinear effects in quantum dots mode-locked comb lasers in Innolume GmbH.

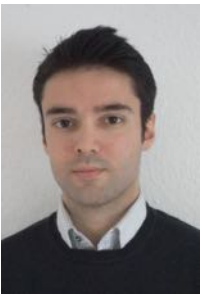

**Sebastian Romero-García** was born in Rute, Spain, in 1984. He received the Ingeniero de Telecomunicación degree from the University of Malaga, Spain, in 2009. From 2009 to 2011, he was a Research Assistant at the Department of Communications Engineering, University of Malaga. Since 2011, he has been with the Institute of Integrated Photonics of RWTH Aachen University, Germany, where he is currently working towards the Ph.D. degree in Electrical Engineering. His research interests focus on the development of integrated photonic devices and systems. Recent publications cover the development of a SiN back-end waveguide technology for biosensing, as well as the development of alignment tolerant couplers for hybrid Silicon Photonics system integration.

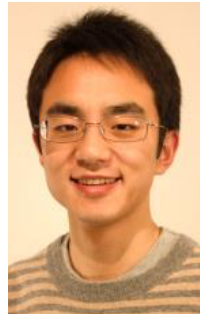

**Bin Shen** was born in Zhejiang, China, in 1986. He received the B.Sc. degree in Material Science and Engineering from the Zhejiang University of Technology, China, with a focus on Polymer Science in 2008 and the M.Sc. in Material Science with a focus on Nanotechnology and Electric Materials from RWTH Aachen University, Germany, in 2012. He joined the Institute of Integrated Photonics of the RWTH Aachen in 2011 as a Master student and is now working towards a Ph.D. degree in Electrical Engineering. After working on test automation methodologies, the focus of his current research is on nonlinear optical devices.

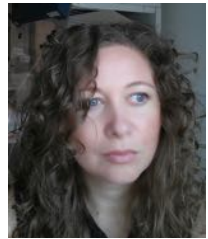

**Anna Sandomirsky** joined Mellanox in 2013 as assembly lab manager. In the advanced technology group, she manages and develops high-end assembly processes for the interconnect products of Mellanox. Prior to joining Mellanox, Anna was an assembly specialist and process engineer at Xloom Communications. She was born in Kharkov, Ukraine, in 1976, and immigrated to Israel in 1997. She graduated from Kharkov University in Computer Science and Patent Law, followed by a B.Sc. in Statistics from the University of Haifa.

**Sylvie Rockman** has been the Sr. Director of Process and Electro-optical Integration at Mellanox Technologies, Israel, since 2013. She is in charge of process technologies and leads a small engineering group responsible for new technologies that is the focal point for micro-fabrication, MEMS and part of the assembly technologies at Mellanox. During this period 2 major R&D laboratories were built including an assembly laboratory introducing high end assembly processes and an electro-optics laboratory for the characterization of laser and photodiodes.

Prior to joining Mellanox Sylvie was Co-Owner of IMH, a turnkey project company in MEMS and related areas. IMH provides comprehensive solutions for high quality projects in the MEMS arena including, project management, product and manufacturing design, engineering follow up on production, upscaling design for high volume production, assembly and packaging.

In 2001 Sylvie joined the founding team of Xloom Communications and served as Senior VP-R&D at Xloom Communications, leading the development of Xloom's opto-transceiver fabrication process. The fabrication process entailed a combination of various MEMS like fabrication steps, as well as several micro-optical assembly steps.

Previous to Xloom Communications, Sylvie was Process Engineering Group Manager at Tower-Semiconductor. Sylvie managed a process engineering team responsible for maintaining various existing processes in Tower's fab-1 as well as for the development of new processes.

Sylvie has extensive knowledge of MEMS fabrication processes, wafer-level assembly processes and deep knowledge of electro-optic components and fiber-optics assembly processes.

In addition, Sylvie is very experienced in leading device and process development projects, including projects that combine MEMS fabrication, wafer-level assembly techniques, electro-optic and optical devices, and various packaging operations. Sylvie holds a B.Sc. in Chemical Engineering and a M.Sc. in Materials Science from the Technion – Israel Institute of technology in Haifa.

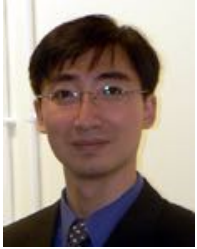

**Chao Li** received his B.S. and M.S. degrees in physics from Tsinghua University, Beijing, China, in 1998 and 2001, respectively. He received his Ph. D. degree in electrical and electronic engineering from The Hong Kong University of Science and Technology, Hong Kong, in 2007. He was a post-doc researcher in the Chinese University of Hong Kong, Hong Kong. He is now a scientist in Institute of Microelectronics, A*STAR, Singapore.

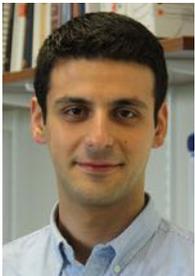

**Saeed Sharif Azadeh** received the M.Sc. degree in Electrical Engineering from Sharif University of Technology, Tehran, Iran, in 2011. His Master thesis focused on the optical properties of HTC superconductor devices. He joined the Institute of Integrated Photonics of RWTH Aachen University, Germany, as a Ph.D. candidate in July 2012. His Ph.D. project includes the simulation and design of Silicon Photonics electro-optical modulators, device fabrication and high frequency device characterization.

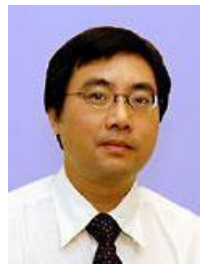

**Guo-Qiang Lo** received the M.S. and Ph.D. degrees in electrical and computer engineering from the University of Texas at Austin, Austin, in 1989 and 1992, respectively. Since 2004, he has been with Institute of Microelectronics, A*STAR, Singapore, where he is currently a Principal Scientist in NanoElectronics/Photonics. He has been managing the silicon-photonic group for the effort of building blocks and integration development.

**Elad Mentovich** received the M.Sc. (Cum Laude) and Ph.D. degrees from the Physical Chemistry Department at Tel Aviv University, Israel. His dissertation is entitled "Realization of the Molecular Transistor Roadmap". He earned his B.Sc. (Cum Laude) in Physics and Materials Engineering at Technion, Haifa. Elad has published extensively in international academic publications such as Applied Physics Letters, The Journal of Nanobiotechnology, Nano Letters and Advanced Materials. Since 2012, he is a Principal Engineer in Process Technologies at Mellanox.

**Florian Merget** received the Dipl.-Ing. and Ph.D. degrees in Electrical Engineering from RWTH Aachen University, Germany, in 2002 and 2008. During his Ph.D., he worked on electrical and optical chalcogenide phase change memory devices and specialized on semiconductor fabrication technologies. In 2008 he joined the Silicon Photonics Group of the Institute of Semiconductor Electronics (led by Prof. H. Kurz) as a project manager. In 2011 he joined the newly established Institute of Integrated Photonics as a Staff-Scientist. His main research interests focus on high-speed data transmission systems and electro-optic components specifically targeted to meet the needs of next generation telecom and datacom applications.

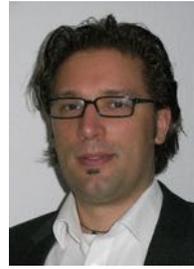

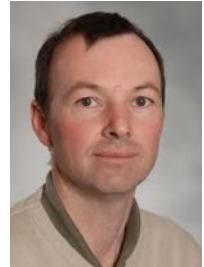

**François Lelarge** received the Diploma in Material Science in 1993 and the Ph.D. degree in 1996, both from the University of Pierre et Marie Curie, Paris, France. From 1993 to 1996 he was with the Laboratory of Microstructures and Microelectronics, CNRS Bagneux, France. His thesis work was devoted to the fabrication and the optical characterization of GaAs/AlAs lateral superlattices grown on vicinal surfaces by MBE. From 1997 to 2000, he was a Postdoctoral Researcher at the Institute of Micro and Optoelectronics, Lausanne, Switzerland. He worked on InGaAs/GaAs quantum wires fabrication by MOCVD regrowth on patterned substrates. Presently, he is in charge of the Epitaxy and New Material Technology team within III-V Lab and coordinator of a project on quantum dash based directly modulated lasers (ANR-DIQDOT). He has published more than 110 papers and communications in national and international conferences, including invited talks and post-deadline papers on quantum dot based devices.

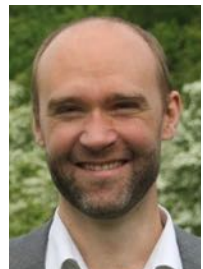

**Jeremy Witzens** received the Ph.D. degree from the California Institute of Technology in 2005, where he worked on Photonic Crystal Micro-Cavities and Photonic Crystal based light steering devices, and his Engineering Diploma from the Ecole Polytechnique, Palaiseau, France. He is currently leading the Institute of Integrated Photonics of RWTH Aachen University, Germany, since 2011. Prior to joining the RWTH, he was a Principal Research Scientist at the University of Washington from 2009 to 2010, a Sr. Staff Engineer at Luxtera Inc. from 2006 to 2009, a Postdoctoral Scholar at the California Institute of Technology in 2005 and an Integrated Optics Designer at Luxtera Inc. from 2002 to 2003. His current research interests include high-performance Silicon Photonics modulators, alignment tolerant optical chip couplers, comb source based communications and sensing systems, group IV light sources, advanced transceiver architectures, hybrid photonic integration and biophotonics. He received a starting independent researcher award from the European Research Council in 2011 and is currently coordinating the European project "Broadband Integrated and Green Photonics Interconnects for High Performance Computing and Enterprise Solutions" (BIG PIPES).